\documentclass{article}

\usepackage[letterpaper, portrait, margin=1in]{geometry}
\usepackage[utf8]{inputenc}
\usepackage{amsmath} 
\usepackage{amssymb}
\usepackage{graphicx}
\usepackage{setspace}
\usepackage{parskip}
\usepackage{array}
\usepackage{booktabs}
\usepackage{hanging}
\usepackage{natbib}
\usepackage{authblk}
\usepackage{url}
\usepackage{orcidlink}

\newcommand{\Grad}{\boldsymbol{\nabla}}
\newcommand{\Div}{\boldsymbol{\nabla} \cdot}
\newcommand{\gvec}{\mathbf{g}}

\newcommand{\dvec}{$\mathbf{d}$ }

\newcommand{\ddt}[1]{\dfrac{d#1}{dt}}

\newcommand{\partialt}[1]{\dfrac{\partial #1}{\partial t}}
\newcommand{\partialx}[1]{\dfrac{\partial #1}{\partial x}}
\newcommand{\partialy}[1]{\dfrac{\partial #1}{\partial y}}

\newcommand{\partialthat}[1]{\dfrac{\partial #1}{\partial \hat{t}}}
\newcommand{\e}[1]{\ensuremath{\times 10^{#1}}}
\usepackage{lineno}

\title{VENUSS: a unified finite-element model of solidifying lava}

\author[a,b,*]{Janine Birnbaum \orcidlink{0000-0003-0873-2989}}
\author[a]{Einat Lev \orcidlink{0000-0002-8174-0558}}
\author[a,c]{Marc W. Spiegelman \orcidlink{0000-0002-5218-7466}}
\author[b]{Jackie E. Kendrick \orcidlink{0000-0001-5106-3587}}
\author[b]{Yan Lavall\'ee \orcidlink{0000-0003-4766-5758}}
\affil[a]{Lamont-Doherty Earth Observatory, Columbia University, NY, USA}
\affil[b]{Ludwig-Maximilian-University Munich, Munich, Germany}
\affil[c]{Department of Applied Physics and Applied Mathematics, Columbia 
University, NY, USA}
\affil[*]{Corresponding author: J.Birnbaum@lmu.de}
\date{}

\singlespace

\providecommand{\keywords}[1]
{
  \small	
  \textbf{\textit{Keywords---}} #1
}

\begin{document}
\maketitle

\begin{abstract}
   The development of a solid rind or carapace at the surface of lava flows and domes results in a transition in deformation mechanism from dominantly viscous to elastic or plastic. This transition has a significant impact on the rate and style of emplacement, including on the construction of channelized flows, over-steepened margins, and flow advance due to lava breakouts. These processes are particularly important in subaqueous, subglacial, and extraterrestrial environments in which cooling is accelerated, requiring models specifically calibrated for these environments. We present a new numerical model, Viscous-Elastic Numerically Unified Solver for Solidifying flows (VENUSS), for cooling and solidifying free surface flows. The model couples a viscous fluid interior with an elastic shell whose thickness grows in response to cooling. As a demonstration of the impact of including a solidified crust in the flow model, we show that a dome-like shape fed from below with an elastic shell coupled to the basal topography results in more lateral expansion and less vertical uplift than a comparable highly-viscous rind, demonstrating the need for lateral transfer of stress in solid layers to accurately interpret and predict dome deformation. 
\end{abstract}

\keywords{Lava flow; Lava dome; Solidification; Fluid-structure interaction; XFEM}

\textbf{\textit{Highlights:}} 
\begin{itemize}
    \item We develop a finite element method for solidifying (a viscous-elastic transition) lava flows and domes.
    \item The model improves our ability to model lava domes and flow under water, ice, and dense atmospheres.
    \item An elastic carapace fed from below deforms differently than a high-viscosity rind, improving hazard assessment.
\end{itemize}

\textbf{\textit{Acknowledgments:}} \par
We acknowledge funding support from the U.S. National Science Foundation (NSF) under awards EAR-1654588 and EAR-1929008 and the European Research Council (ERC) under Magma Outgassing During Eruptions and Geothermal Exploration (MODERATE No.101001065). We would additionally like to thank Hannah Dietterich for constructive discussion which improved the manuscript. 

\newpage

\section{Introduction}
Effusive volcanic eruptions such as lava flows and domes are common and can pose a significant hazard to life and property. Lava flows can travel for many kilometers and destroy buildings and infrastructure. Recent costly examples include the 2018 eruption of Kīlauea volcano that resulted in financial loss at an estimated USD\$500M and the 2021 eruption of Cumbre Vieja at an estimated loss of USD\$1B \citep{EMDAT}. While lava domes are typically restricted to the near-vent region, collapses of the over-steepened sides or over-topping of confining topography can generate pyroclastic density currents even when eruption rates are not significantly heightened or effusion has ceased \citep{Rose1976,Calder2002,Carr2022}. Accurate forecasting of eruption hazards requires physics-based models that account for the driving processes of transport and emplacement. \par

The emplacement of effusive volcanic products is subject to a variety of internal magmatic processes such as changes in lava rheology and texture, the rate and steadiness of magma supply, and external controls including pre-existing topography and loss of heat to the surrounding environment. Melt viscosity is highly temperature dependent, spanning many orders of magnitude \citep[e.g.,][]{Giordano2008} between initial eruption temperatures and ambient environmental conditions. The rheology of  magmatic suspensions, comprising silicate melt and suspended crystals and bubble cargoes is a complex function of melt viscosity and the interaction with suspended phases \citep[e.g.,][and references therein]{Caricchi2007,Lavallee2007,Costa2009,Pistone2012,Kolzenburg2022}. As lavas and magmas cool below a threshold temperature region they solidify via 1) crystallization, which reduces the amount of melt available for flow, causing corresponding changes in melt chemistry, and 2) the viscosity of the melt increases until the viscous relaxation timescale becomes long with respect to the deformation timescale, causing solid-like behavior \citep{Dingwell1989}; the threshold temperature of this transition, the glass transition, depends on lava chemistry and strain rate \citep{Ryan1981}, and is met at lower bulk strain rates in crystal-bearing suspensions due to localization of deformation between crystals \citep{Cordonnier2012,Coats2018,Vasseur2023}. Observations from natural systems \citep{Chadwick2013,Darmawan2017,Grosfils2000,Pinkerton1994,Stofan2000,Tuffen2013}, laboratory experiments \citep{Cashman2006,Fink1993,Fink1995,Fink1998,Gregg1995,Kerr2006,Lev2019,Rader2017,Peters2022}, and mathematical models \citep{Crisp1990,Iverson1990,Quick2016,Quick2017}, highlight the important role of solidification and crust fracture and disruption in controlling the runout and morphology of lava flows and domes. Additionally, the growth of a coherent crust is a dominant process in controlling subaqueous \citep{Chadwick2013,Perfit1998,Tan2016}, subglacial \citep{Russell2014}, and extraterrestrial eruptions into dense atmospheres \citep[e.g.,][]{Fink1998,Stofan2000} which contributes to the lack of lava flow models specifically calibrated for these environments. \par

\begin{figure}
\begin{center}
\includegraphics[width=\textwidth]{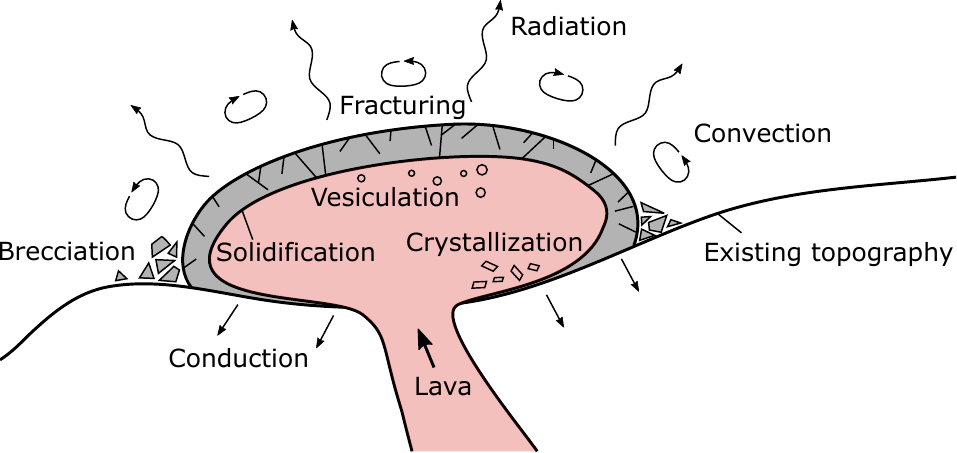}\\
\caption{Schematic of a lava dome with a solidified and fractured shell.}
\label{fig:dome-schematic}
\end{center}
\end{figure}

Solidification plays a key role in controlling the morphology of high-viscosity lava flows and lava domes. These effusive eruptions are characterized by high aspect ratios of height to areal extent, and often feature solidified and highly fractured surfaces and flow margins \citep{Andrews2021,Fink1995,Fink1998,Stofan2000}. Aspect ratio and fracturing are controlled by rate and steadiness of magma supply, cooling rate, and strength of the solidified material \citep{Cashman2006,Fink1993,Fink1995,Fink1998,Lev2019,Iverson1990,Peters2022}. Fractures create permeable pathways for outgassing that may be important in outgassing the interior of viscous lava flows and domes, and can contribute to simultaneous effusive and explosive hazards \citep[e.g.,][]{Lavallee2013,Castro2014,Crozier2022,Kendrick2016}.
\par

In unconfined lava flows, solidifying margins channelize the lava, and, if crusted over, can insulate lava flowing in tubes and promote longer runout distances \citep{Crisp1990,Pinkerton1994}. Eventually, solidification of the flow front arrests flow, behind which continued lava supply can lead to flow inflation and lava flow breakouts that restart previously stalled portions of the flow field. In some cases, lateral lava flow breakouts can permanently change the primary lava pathway, such as during the 2014-2015 Pāhoa flow at Kīlauea \citep{Poland2016}, or account for much of the late-stage flow propagation as in the case of the 2011-2012 eruption of Puyehue-Cord\'on Caulle \citep{Tuffen2013}. Models that can predict the location, timing, and evolution of lava flow breakouts remain an outstanding challenge despite their direct hazard implications.  \par 

Numerical models allow us to make quantitative predictions about the modes and extent of lava flow and dome emplacement, and determine the effect of varying system parameters such as magma viscosity (highly dependent on temperature and physical and chemical composition), mass flux and flux rate, and system geometry. Scaling between analogue experiments and natural systems is not always straightforward, and it is usually difficult to match more than one dimensionless ratio in the system. Numerical models are instrumental in bridging the different temporal and spatial scales to integrate field observations with controlled experiments to interpret emplacement conditions and predict future eruptions.

\section{Existing models}
\subsection{Volcanological community models}
Numerical simulation of lava flows is an area of active research that encompasses a wide range of techniques and complexities. The simplest of these models exploit controlling factors such as underlying slope and topography to predict lava flow paths without consideration of material properties or eruption conditions. This technique includes calculating paths of steepest descent such as Downflow \citep{Favalli2005} or LASZLO \citep{Bonne2008}, or inundation using cellular automaton models, for example MOLASSES \citep{Connor2012}, MAGFLOW \citep{DelNegro2004}, or SCIARA \citep{Barca2004}. \par
\label{section:community_codes}

Some existing models incorporate simplified solutions for momentum that solve one- or two-dimensional depth-averaged versions of the shallow water equations, for example FLOWGO \citep{Harris2001}, VOLCFLOW \citep{Kelfoun2015}, and Lava2D \citep{Hyman2022}. Such models may include additional treatment of non-Newtonian (e.g., Bingham-plastic) rheology or a temperature-viscosity structure with depth \citep[e.g.][]{Crisp1990,Hyman2022}. Even with an increase in viscosity towards the surface, the presence of a mechanically coherent crust is rarely considered, or is treated using a simplified approach. Treatment of a solidified crust has primarily been accounted for in the thermal budget, in which thermal insulation provided by a crust may be parameterized \citep[e.g.][]{Harris2001}, or treated as a two-layer model \citep[LavaSIM, ][]{Fujita2016}. The effect of a crust on the momentum equations is only considered in the depth-integrated approach \citep{Iverson1990,Quick2016,Quick2017}. The advantage of depth-averaged or simple spreading models is their low computational cost that allows for rapid response to unrest or eruption, and probabilistic hazard forecasting using large ensembles. However, they must be calibrated to natural and analogue observations which are biased toward subaerial eruptions, typically of low to intermediate viscosity \citep{Cordonnier2016,Dietterich2018}, and they generally do not support restarting of a stalled flow through lava flow breakouts. To simulate breakouts, the above models may be run using the flow margin as a new source of lava, which requires \emph{a priori} knowledge of the location and timing of a breakout. \par 

In hazard assessment, breakouts can be simulated through a probabilistic framework \citep[e.g., MrLavaLoba, ][]{deMichieliVitturi2017}. The process of fracture and breakouts leads to strong feedbacks in the localization of flow. In the case where the entire crust or levee is close to the yield strength, which may often be the case at flow margins, small-scale details of the crust geometry or material heterogeneity may control the fracture location in difficult-to-predict ways. This leads to  solutions in the evolution of flow fields, which requires a probabilistic or ensemble-type model to assess hazard. However, a physics-based estimate of the likelihood of failure allows for a deeper understanding of the processes that lead to breakouts and may help identify potential precursors, which remains an outstanding challenge for the volcanological community. \par

\subsection{General flow models}
Beyond lava dynamics, free-surface flows, fluid-solid interactions, solidification, and fracture initiation and propagation have a variety of engineering, medical, physical science, and industrial food and drug manufacturing applications, and have thus been studied previously using numerical techniques. In particular, fully three-dimensional models using finite element or finite volume approaches allows for solving a wide range of general problems. There are numerous commercial (e.g., FLOW-3D, COMSOL, ANSYS FLUENT) and open-source (e.g., Open-FOAM, \citealp{Jasak2007}; ELMER) software capable of multi-physics modeling of flows. Two- and three-dimensional models of lava flows with coupled temperature and momentum fields are computationally expensive, and are thus more suitable for building understanding and interpretation of experiments and individual flows than for probabilistic hazard assessment. The combination of level set methods and the extended finite element method (XFEM) improves the capability of capturing behavior around interfaces and singularities without the need for extensive remeshing. This technique has been previously applied to two-fluid interfaces \citep[e.g.,][]{Chessa2003,Chessa2003b,Fries2009,GroS2007}, fluid-structure interaction \citep[][]{Jenkins2015,Kreissel2012,Mayer2010}, and solidification \citep[e.g.,][]{Chessa2002}.

\subsection{Contributions of this model}
Below, we develop a finite element model Viscous-Elastic Numerically Unified Solver for Solidifying flows (VENUSS). The model uses the GetFEM C++ library \citep{Renard2020} and Python interface. The model couples a cooling, free-surface viscous flow with a solidifying shell that specifically addresses the challenges of modeling lava flows and domes in two-dimensional planar or axisymmetric geometries. This model allows for the calculation of stresses in the elastic shell for evaluation of the likelihood of fracture, a key step in allowing us to move towards physics-based predictions of the conditions and timing of lava flow breakouts. 

\section{Methods}
\subsection{Governing equations}
We consider a continuous material with either viscous or elastic properties that satisfies conservation of mass: 
\begin{equation}
    \label{eq:mass_conservation}
    \partialt{\rho} + \Grad \cdot \left( \rho \mathbf{u} \right) = 0 \: , 
\end{equation}
where $\rho$ is the density of the material, $\mathbf{u}$ is the velocity, and $t$ is time. All symbols and variables are listed in Table \ref{tab:variables}. The code has an option for the material to be nearly incompressible: 
\begin{equation}
    \Div \mathbf{u} \approx \varepsilon p \: , 
\end{equation}
or linearly compressible: 
\begin{equation}
    \beta = \frac{1}{\rho}\frac{\partial \rho}{\partial p} \: ,
\end{equation}
where $\beta$ is the compressibility of the material and $p$ is pressure. Which results in a constitutive relationship for density: 
\begin{equation}
    \rho = \rho_{0} \exp (\beta (p - p_0)) \: , 
\end{equation}
where $\rho_0$ is the reference density, and $p_0$  the reference pressure at ambient conditions. We linearize the density dependence on pressure according to: 
\begin{equation}
    \rho \approx \rho_0 (1 + \beta (p-p_0)) \: .
\end{equation}
Substitution into the mass conservation equation (eq. \ref{eq:mass_conservation}) leads to: 
\begin{equation}
    \beta \partialt{p} + \beta \mathbf{u} \cdot \Grad p + \left(1 + \beta(p-p_0) \right) \Div \mathbf{u}  = 0\: .
\end{equation}
\par

Conservation of momentum in the viscous field is given by the linearized momentum conservation:
\begin{equation}
    \rho \partialt{\mathbf{u}} = \Div \boldsymbol{\sigma} - \Grad p + \rho \mathbf{g} \: 
\end{equation}
where $\gvec$ is acceleration due to gravity and $\boldsymbol{\sigma}$ is the deviatoric stress tensor which relates to the strain-rate tensor, $\boldsymbol{\dot\varepsilon}$: 
\begin{equation}
    \boldsymbol{\dot\varepsilon} = \frac 1 2 (\Grad \mathbf{u} + (\Grad \mathbf{u})^T) \: , 
\end{equation}
through: 
\begin{equation}
    \boldsymbol{\sigma} = 2 \mu^* \boldsymbol{\dot\varepsilon} + \lambda^* tr( \boldsymbol{\dot\varepsilon}) \mathbf{I} \: ,
\end{equation}
where $\mu^*=\eta$ is the effective dynamic viscosity and $\lambda^* = (\zeta - \frac 2 3 \eta )$. We make the Stokes approximation in which the bulk viscosity $\zeta = 0$. We provide the option to set the acceleration to zero, reducing the problem to the steady Stokes equation. \par 
Momentum in the elastic region is given by:  
\begin{subequations}
\begin{align}
    \rho \partialt{\mathbf{u}} &=  \Div \boldsymbol{\sigma} + \rho \mathbf{g} \: , \\
    \boldsymbol{\sigma} &= 2G\boldsymbol{\varepsilon} + (\lambda + G) tr(\boldsymbol{\varepsilon}) \mathbf{I} \: ,  \\
    \boldsymbol{\varepsilon} &= \frac 1 2 (\Grad \mathbf{d} + (\Grad \mathbf{d})^T) \: , 
\end{align}
\end{subequations}
where $\boldsymbol\varepsilon$ is the elastic strain tensor, $G$ is the shear modulus, $\lambda$ is Lam\'e's first parameter, and \dvec is the elastic displacement. \par
The strain displacement field in an Eulerian framework evolves according to: 
\begin{equation}
    \partialt{\mathbf{d}} = \mathbf{u_n} - \mathbf{u_n} \cdot \Grad \mathbf{d_n}
\end{equation}
We approximate the strain displacement field as a function of previous states of the system and the velocity according to the time discretization scheme:
\begin{equation}
    \alpha_0 \mathbf{d}_{n} = \Delta t (\mathbf{u} - \mathbf{u} \cdot \Grad \mathbf{d}_{n}) - \alpha_1 \mathbf{d}_{n-1} - \alpha_2 \mathbf{d}_{n-2} - ... \, 
\end{equation}
where $\Delta t$ is the time step and $\alpha_n$ are coefficients for the backward differentiation formula time integration scheme. Substituting in to the momentum equation for the elastic region:
\begin{subequations}
\begin{align}
    \rho \partialt{\mathbf{u}} &=  \Grad \cdot \boldsymbol{\sigma} - \Grad \cdot \boldsymbol{\sigma_0} + \rho \mathbf{g} \: , \\
    \boldsymbol{\sigma_0} &= 2G\boldsymbol{\varepsilon_0} + (\lambda + G) tr(\boldsymbol{\varepsilon_0}) \mathbf{I} \: , \\
    \boldsymbol{\varepsilon_0} &=\frac 1 2( \Grad(\frac{\alpha_1}{\alpha_0} \mathbf{d}_{n-1} + \frac{\alpha_2}{\alpha_0} \mathbf{d}_{n-2} + ... + \frac{\Delta t}{\alpha_0} \mathbf{u} \cdot \Grad \mathbf{d}_{n-1}) +  (\Grad(\frac{\alpha_1}{\alpha_0} \mathbf{d}_{n-1} + \frac{\alpha_2}{\alpha_0} \mathbf{d}_{n-2} + ... + \frac{\Delta t}{\alpha_0} \mathbf{u} \cdot \Grad \mathbf{d}_{n-1}))^T)\: , 
\end{align}
\end{subequations} 
We now combine the viscous and elastic cases to create a unified model after \citet{Bordere2014} and \citet{Hubner2004}: 
\begin{equation}
    \rho \partialt{\mathbf{u}} =  \Div  2 \mu^* \boldsymbol{\dot\varepsilon} + \lambda^* tr( \boldsymbol{\dot\varepsilon}) \mathbf{I}  - \Grad \cdot \boldsymbol{\sigma_0} + \rho \mathbf{g}
\end{equation}
where in the fluid region $\mu^*=\eta$ is the dynamic viscosity and $\lambda^* = -\frac 2 3 \eta $, and the residual stress tensor $\sigma_0$ = 0, and in the solid region, $\mu^* = \frac{\Delta t}{\alpha_0} G$ and $\lambda^* = \frac{\Delta t}{\alpha_0} \lambda$.

\par

Conservation of energy (heat) is given by: 
\begin{equation}
    \rho c_p \partialt{T} = \Grad \cdot (k \Grad T) - \rho c_p \mathbf{u} \cdot \Grad T + q_T \: , 
\end{equation}
where $T$ is the temperature, $k$ is the thermal conductivity, $c_p$ is the specific heat capacity, and $q_T$ is a source/sink term that could include temperature lost by radiation or forced convection in the air. \par

Viscosity $\eta$ is a function of melt viscosity $\mu$, which in turn depends on temperature and lava chemistry according to the Vogel–Fulcher–Tammann (VFT) equation, and the contribution of suspended phases $\eta_r$: 
\begin{subequations}
\begin{align}
    \mu &= \exp \left( A + \frac{B}{T-C} \right) \: , \\
    \eta &= \mu \eta_r \: , 
\end{align}
\end{subequations}
where $A$, $B$, and $C$ are empirical parameters \citep{Giordano2008}, and $\eta_r$ is a function of the suspended phases \citep{Mader2013,Costa2009,Birnbaum2021}.

\begin{table}[h!]
\fontsize{10}{12}\selectfont
\centering
\begin{tabular}{|c|l|c|}
    \hline
    Variable & Description & Units \\
    \hline
    $x$, $y$ & Coordinate directions & \\
    $t$ & Time & s \\ 
    $\mathbf{u}$ & Velocity & m/s \\
    $p$ & Pressure & Pa \\
    $p_0$ & Ambient pressure & Pa \\
    $\boldsymbol{\sigma}$ & Viscous/elastic stress tensor & Pa \\
    $\boldsymbol{\sigma_0}$ & Residual elastic stress tensor & Pa \\
    $\boldsymbol{\dot\varepsilon}$ & Strain-rate tensor & 1/s \\
    $\boldsymbol{\varepsilon}$ & Strain tensor & m/m \\
    $\mathbf{d}$ & Elastic displacement & m \\
    $T$ & Temperature & $^\circ$C \\
    $q_T$ & Heat source/sink term & $^\circ$C/s \\ 
    $\Psi$ & Free-surface level-set & m \\
    $\Psi_{smoothed}$ & Smoothed free-surface level-set & m \\
    $F$ & Extension velocity & m/s \\
    $F_{stabilized}$ & Stabilized extension velocity & m/s \\
    
    
    $\rho$ & Density & kg/m$^3$ \\
    $\rho_0$ & Reference density at $p_0$ & kg/m$^3$ \\
    $\beta$ & Compressibility & 1/Pa \\
    $\mathbf{g}$ & Gravitational acceleration & m/s$^2$ \\
    $\lambda^*$ & Unified viscous-elastic Lam\'e's first parameter & \\
    $\mu^*$ & Unified viscous-elastic Lam\'e's second parameter & \\
    $\mu$ & Liquid viscosity & Pas \\
    $\eta$ & Effective viscosity & Pas \\
    $\eta_r$ & Relative viscosity & Pa/Pa \\
    $\zeta$ & Bulk viscosity & Pas \\
    $G$ & Shear Modulus & Pa \\
    $\lambda$ & Linear elastic Lam\'e's second parameter & Pa \\
    $k$ & Thermal conductivity & W/m$\cdot$K \\
    $A$, $B$, $C$ & Vogel-Fulcher-Tammann parameters & \\
    $\alpha_0$, $\alpha_1$, $\alpha_2$ & Backward differentiation formula (BDF) coefficients & \\
    $\epsilon$ & Pseudo-viscosity stabilization parameter & \\
    $\kappa_\Psi$ & Pseudo-diffusivity stabilization parameter & \\
    $K$ & Curvature of level set &  \\
    $K_{\text{mean}}$ & Mean curvature of level set & \\

    \hline
\end{tabular}
\caption{List of variables in the conservation of mass, momentum, energy, and displacement and level-set update equations.}
\label{tab:variables}
\end{table}

\subsection{Scaling analysis}
From the structure of the momentum and energy equations, we identify various time and length scales. We scale according to $t = \tau\hat{t}$, $x = L\hat{x}$, $y = L\hat{y}$, $u = U\hat{u}$, $d = U\tau\hat{d}$, $T = \Theta \hat{T}$, $\rho = \rho\hat{\rho}$, and $g = g\hat{g}$ (Table \ref{tab:scaling}). Substituting into the equation for momentum in the viscous domain and rearranging gives: 
\begin{equation}
    \frac{1}{\tau} \hat{\rho} \partialthat{\mathbf{\hat{u}}} =  \frac{\eta}{\rho L^2}\Grad \cdot \boldsymbol{\hat{\dot\sigma}} + \frac{g}{U} \hat{\rho} \mathbf{\hat{g}} \: , 
\end{equation}
which has possible timescales dominated by advection, $\tau_{\text{adv}}=\frac{L}{U}$, viscous relaxation, $\tau_{\text{visc}}=\frac{\rho L^2}{\eta}$, and gravity, $\tau_{\text{grav}}=\sqrt{\frac{L}{g}}$. \par 
Substituting into the equation for momentum in the elastic domain and rearranging gives: 
\begin{equation}
    \frac{1}{\tau} \hat{\rho} \partialthat{\mathbf{\hat{u}}} =  \left(\frac{v_p}{L}\right)^2 \tau \Grad \cdot \boldsymbol{\hat{\sigma}} - \left(\frac{v_p}{L}\right)^2 \tau \Grad \cdot \boldsymbol{\hat{\sigma_0}}+ \frac{g}{U} \hat{\rho} \mathbf{\hat{g}} \: , 
\end{equation}
which adds an additional timescale for elasticity, $\tau_{\text{elas}}=\frac{L}{v_p}$, where $v_p$ is the compressional wave velocity. \par 
From the energy equation, we arrive at: 
\begin{equation}
    \frac{1}{\tau}\partialthat{\hat{T}} = \frac{\kappa}{L^2}\Grad \cdot (\Grad \hat{T}) - \frac{U}{L}\mathbf{u} \cdot \Grad T + \frac{1}{\Theta}q_T \: , 
\end{equation}
which adds a timescale for diffusion, $\tau_{\text{diff}} = \frac{L^2}{\kappa}$. \par
The ratios of these timescales to the advective timescale yields a set of well-known non-dimensional numbers:
\begin{subequations}
    \begin{alignat}{2}
        \frac{\tau_{\text{visc}}}{\tau_{\text{adv}}} &= \mathrm{Re} & &= \frac{\rho L U}{\eta} \: , \\
        \frac{\tau_{\text{grav}}}{\tau_{\text{adv}}} &= \mathrm{Fr} & &= \frac{U}{\sqrt{g L}} \: , \\
        \frac{\tau_{\text{elas}}}{\tau_{\text{adv}}} &= \mathrm{Ma} & &= \frac{U}{v_p} \: , \\
        \frac{\tau_{\text{diff}}}{\tau_{\text{adv}}} &= \mathrm{Pe} & &= \frac{L U}{ \kappa} \: ,
    \end{alignat}
\end{subequations}
where $\mathrm{Re}$ is the Reynolds number, $\mathrm{Fr}$, the Froude number, $\mathrm{Ma}$, the Mach number, and $\mathrm{Pe}$, the P\'eclet number. A summary of the scaling variables and timescales is given in Table \ref{tab:scaling}. \par

Lava and magma transport can span large ranges of the above non-dimensional ratios, in part due to the large range of viscosities and velocities found in natural flows. Typical magma and lava melt densities fall between 2500-2800 kg/m$^3$, and suspension densities can be increased slightly by crystallization, or decreased by the presence of vapor bubbles, to span a typical range of 750-3300 kg/m$^3$, although variation within a single flow is usually smaller. Viscosities vary from as low as $\sim$50 Pas up to the glass transition, usually defined at 10$^{12}$ Pas. Thermal diffusivity is typically in the range 10$^{-6}$ - 10$^{-7}$ m$^2$/s. Acoustic velocity of magmas lie between 3.5-5 km/s. Velocity and length scales depend greatly on the system geometry. Restricting the discussion to lava flows and lava domes, thicknesses may vary between tens of centimeters to hundreds of meters, and lateral distances up to several kilometers. Velocities can be as high as tens of meters per second, although centimeters to meters per second are more typical advance rates \citep{Harris2015}. As a result, the Reynolds number for lava flows are typically low ($\lesssim$300), the Froude number is typically low ($\ll$ 1), but can be high ($\lesssim$5) \citep{Dietterich2022}, the Mach number is always very low ($\lesssim$10$^{-2}$), and the P\'eclet number is very high ($\gtrsim$10$^{4}$) at the flow scale, but diffusion is important in the cooling of thin crusts and skins. Therefore, in this work we use a linearized momentum equation (low $\mathrm{Re}$ limit), and neglect the propagation of acoustic waves (low $\mathrm{Ma}$ limit).

\begin{table}[h!]
\fontsize{10}{12}\selectfont
\centering
\begin{tabular}{|c|l|c|}
    \hline
    Variable & Description & Units\\
    \hline
    $\tau$ & Time & s \\
    $L$ & Length & m \\
    $U$ & Velocity & m/s \\
    $\Theta$ & Temperature & K \\
    $\rho$ & Density & kg/m$^3$ \\
    $g$ & Gravitational acceleration & m/s$^2$ \\
    $v_p$ & Compressional wave velocity & m/s \\

    \hline
    Timescales & Description & Value \\
    \hline
    $\tau_{\text{adv}}$ & Advection & $\frac{L}{U}$ \\
    $\tau_{\text{visc}}$ & Viscous relaxation & $\frac{\rho L^2}{\eta}$ \\ 
    $\tau_{\text{grav}}$ & Gravitational relaxation & $\sqrt{\frac{L}{g}}$ \\
    $\tau_{\text{elas}}$ & Elastic propagation & $\frac{L}{v_p}$ \\
    $\tau_{\text{diff}}$ & Diffusion & $\frac{L^2}{\kappa}$ \\
    
    \hline
\end{tabular}
\caption{List of dimensional scales.}
\label{tab:scaling}
\end{table}

\subsection{Weak form of the equations}
We find the weak form of the equations by multiplying by a trial function in $p$, $p_t$, on the conservation equation: 
\begin{equation}
    \int p_t \ddt{\rho} d\Omega + \int p_t \Div (\rho u) d\Omega = 0 \: ,
\end{equation}
a trial function in $u$, $u_t$, on the momentum equation: 
\begin{equation}
    \int \mathbf{u}_t \cdot \ddt{\rho \mathbf{u}} d\Omega - \int \Grad \mathbf{u}_t \cdot (\boldsymbol{\sigma} - \boldsymbol{\sigma}_0) d\Omega - \int \mathbf{u}_t \cdot \rho \mathbf{g} d\Omega = \oint (\mathbf{u}_t \cdot (\boldsymbol{\sigma} - \boldsymbol{\sigma}_0)) \cdot \mathbf{\hat{n}} d\Gamma \: , 
\end{equation}
and a trial function in $T$, $T_t$, on the heat equation: 
\begin{equation}
    \int \rho c_p T_t\ddt{T} d\Omega + \int \rho c_p T_t (\mathbf{u} \cdot \Grad T) d\Omega  + \int \Grad T_t (k \Grad T) d\Omega - \int T_t q_T d\Omega = \oint (T_t k \Grad T) \cdot \mathbf{\hat{n}} d\Gamma \: .
\end{equation}

\subsection{Representation of the interfaces with level sets}
\label{section:level-sets}
The level set method is a common technique for representing interfaces within the computational domain \citep{Adalsteinsson1995,Sethian2003}. We utilize up to three independent level sets. The first level set represents the interface between two immiscible fluids that can have different density, compressibility, viscosity, and thermal diffusivity. The fluid (lava/magma) below the level set, always fluid 1, can have a viscosity that depends on temperature and up to three parameters \citep[e.g. the VFT equation;][]{Richet1996,Dingwell1998}. Fluid 2 above the level set is a fluid environment that could be water or air, although a logical future extension would be to add an additional level set in this fluid 2 region that could handle the phase transition between ice and water. A second level set is defined along an isothermal contour at the glass transition temperature and marks the boundary between the viscous and elastic portions of the domain. In each time step the level set is updated based on the temperature field and is not, in general, restricted to being the signed distance function. The third level set is the basal topography along which no-slip or no-normal flow conditions may be applied.

The fluid interface level set (level set 1), $\Psi$, is advected using the velocity solution on the zeroth contour of the level set, F. The velocity is stabilized by an imposed pseudo-viscosity that prevents sharp corners from forming on the interface such that: 
\begin{equation}
    \mathbf{F_{\text{stabilized}}} = \mathbf{F}  - \frac{\epsilon}{\eta} (K - K_{\text{mean}}) \frac{\Grad \Psi}{|\Grad \Psi|} \: , 
    \label{eq:stabilized_velocity}
\end{equation}
where $\epsilon$ is a stabilizing factor that is much smaller than 1. $K$ is the curvature of the interface and $K_{mean}$ is the mean curvature of the interface \citep{Sethian1997}. We threshold the maximum value of the smoothing correction according to the maximum stable velocity ($\Delta x / 2 \Delta t$) to prevent instability in regions of high curvature. Curvature and the normal to the interface are calculated using second-order finite-differences on a centered 9-point stencil on a smoothed level set: 
\begin{equation}
    \Psi_{\text{smooth}} + \kappa_{\Psi} \Grad^2 \Psi_{\text{smooth}} = \Psi \: , 
\end{equation}
which acts to remove high-frequency signal in the curvature calculation. \par

The stabilized velocity of the interface, $F_{\text{stabilized}}$, is extended throughout the model domain from the interface location using the fast marching method ($F_{\text{ext}}$). This results in a velocity field for the level sets that is the fluid velocity on the interface, and satisfies: 
\begin{equation}
    \Grad |\mathbf{F}_{\text{ext}}| \cdot \Grad \Psi=0 \: .
\end{equation}
We propagate the level set according to:
\begin{equation}
    \partialt{\Psi} + \mathbf{F}_{\text{ext}} \cdot \Grad \Psi_{\text{smooth}} = 0 \: , 
\end{equation}
which has a weak form: 
\begin{equation}
    \int \Psi_t \partialt{\Psi} d\Omega  + \int \Psi_t \mathbf{F}_{\text{ext}} \cdot \Grad \Psi_{\text{smooth}} d\Omega = 0 \: .
\end{equation}
This method enforces the level set to remain the signed distance function everywhere, eliminating the need for re-initialization of the level set to prevent bunching up or stretching out of the level sets with time \citep{Adalsteinsson1995,Sethian1997,Sethian2003}. We enforce a volume conservation by iterating over the change in area before and after propagation, accounting for any volume source or sink terms, and distribute any change in volume equally to the interface above any basal topography \citep[e.g.,][]{vanderPijl2008}. This volume correction is not performed when the interface between the two fluids intersects a domain boundary. \par

The second level set is updated by solving the temperature field and setting the zero contour to the location of the glass transition temperature. Level set two does not necessarily remain the signed distance function. The third level set is not updated through time (the basal topography does not change). \par

\subsection{Elastic displacement}
Stress in the solidified material is approximated at each time step according to the velocity field and the previous state of stress. After the calculation of the instantaneous velocity field, the elastic displacement is updated according to: 
\begin{equation}
    \ddt{\mathbf{d}} + \mathbf{u} \cdot \Grad \mathbf{d} = \mathbf{u} \: , 
\end{equation}
which has a weak form: 
\begin{equation}
    \int \mathbf{d}_t \cdot \ddt{\mathbf{d}} d\Omega + \int \mathbf{d}_t \cdot (\mathbf{u} \cdot \Grad \mathbf{d}) d\Omega = \int \mathbf{d}_t \cdot \mathbf{u} d\Omega \: .
\end{equation}
\par
In the unsolidified regions, we do not allow for the development of elastic stress, and therefore fix the elastic displacement field at zero, determined by the location of the updated level set field. Additionally, upon progressive solidification, we assume that newly solidified material has no elastic stress; given the choice of constitutive relationship, in which elastic stress is proportional to elastic strain, we satisfy this condition by assuming newly solidified material has no elastic strain, that is, there is no gradient in the displacement field. As a result, material that is newly solidified has the same elastic displacement as the nearby previous surface, which is propagated away from the surface using the fast marching method. Regions of low temperature with a size below a minimum threshold of 9 connected nodes are treated as viscous. \par

\subsection{Meshing and basis functions}
\begin{figure}
\begin{center}
\includegraphics[width=4in]{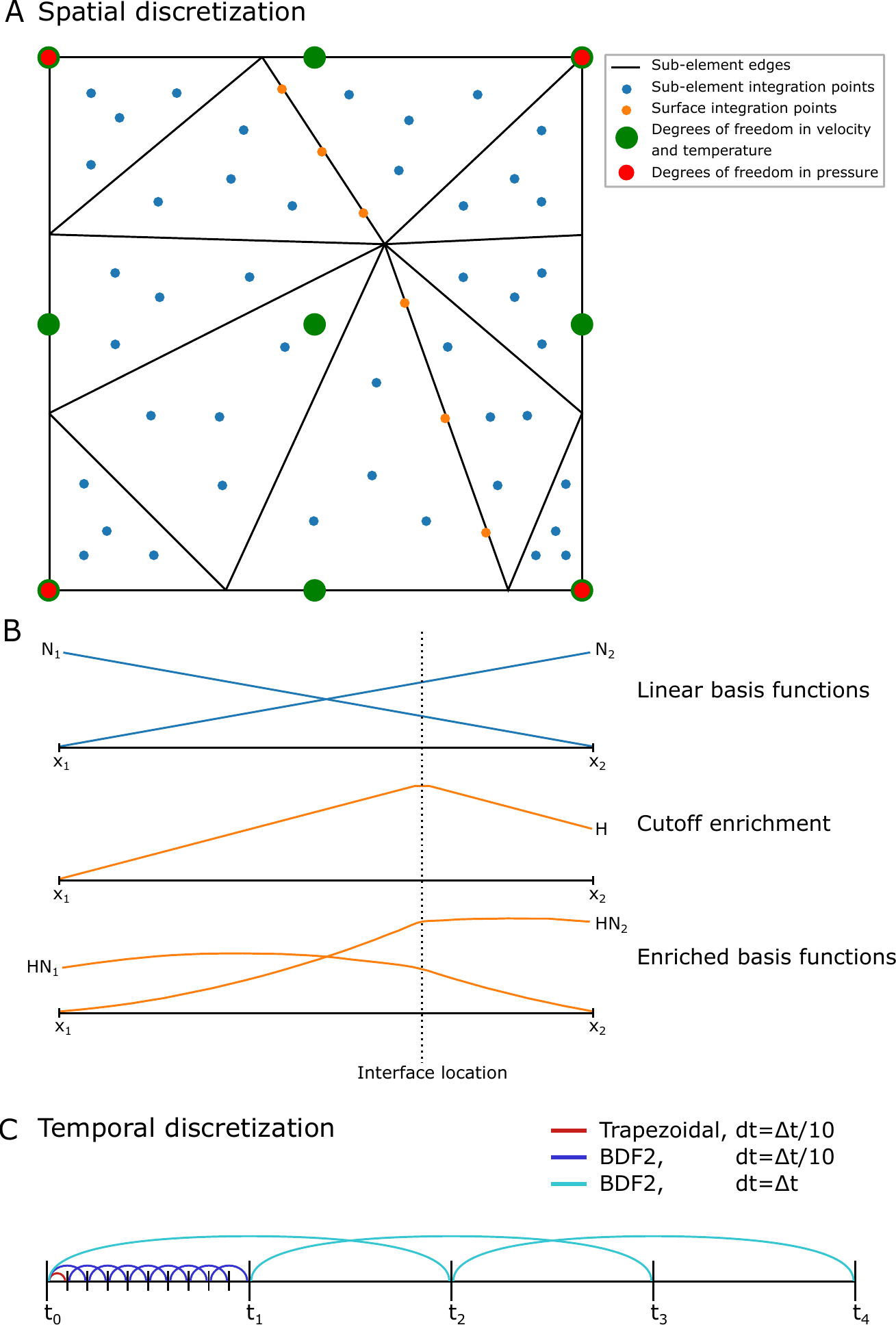}\\
\caption{A) Example sub-divided element cut by the free surface. Blue markers indicate the location of integration points in each sub-element. Orange markers indicate the location of integration points on edges of the free surface. Green and red markers indicate locations of the degrees of freedom in pressure and velocity respectively. B) Polynomial basis functions (N$_i$), cutoff enrichment (H), and their product, the enriched basis functions (HN$_i$). C) Visualization of time-stepping scheme showing the initial step divided into ten sub-steps using the trapezoidal rule for the first sub-step (red), backward differentiation formula (BDF) 2 for the remaining sub-steps (dark blue), and BDF2 for remaining all other time steps (cyan).}
\label{fig:integration_points}
\end{center}
\end{figure}

We generate a uniform rectilinear grid cut by up to three level sets. We use continuous quadratic basis functions on the velocity field and continuous linear basis functions in pressure (Q2-Q1). We use continuous quadratic basis functions on the temperature field. The basis functions on the pressure and material property (density, viscosity, thermal diffusivity) fields are enriched (XFEM) at the free-surface and topography level sets with a Heaviside jump which adds an additional degree of freedom that quantifies a discontinuity within the cell, aligned with the level set, which allows for step-changes in pressure across the interface. The temperature field can either be unenriched, enriched with a cutoff function (Fig. \ref{fig:integration_points}B) that allows for higher-frequency changes near the level set \citep{Chahine2008}, or enriched with a Heaviside jump. In elements cut by one or more level sets, the mesh is decomposed into triangular sub-elements for integration (Fig. \ref{fig:integration_points}A) according to \citet{Bank1983}. The level set representation is linear within each element.

\subsection{Temporal discretization}
We use finite differences in time with a constant time step. We use the backward differentiation formula (BDF), which is an implicit, linear multi-step method, via a two-step formula, BDF2:
\begin{equation}
    \frac 3 2 y_n = \Delta t f(t_n,y_n) + 2 y_{n-1} - \frac 1 2 y_{n-2} \: , 
\end{equation}
where $\Delta t$ is the time step and $f(t_n,y_n)$ is the derivative with respect to time. The first time step is divided into ten smaller sub-steps ($\Delta t^* = \Delta t/10$) to avoid introducing a higher-order error term in the first time step which typically also has the highest frequency solution (especially for temperature). We bootstrap in to the method for the first sub-step using the trapezoidal rule: 
\begin{equation}
    2 y_n = \Delta t^* ( f(t_n,y_n) + f(t_{n-1},y_{n-1}) ) + 2 y_{n-1} \: ,
\end{equation}
followed by nine sub-steps using BDF2 to find the solution at $t=\Delta t$ (Fig. \ref{fig:integration_points}C). The initial condition in pressure and velocity is optionally either (1) ambient pressure and at rest everywhere, or (2) from the solution to steady incompressible stokes flow. In the case of a solidified crust, the elastic displacements are initially zero everywhere. The initial temperature field is prescribed. \par

\begin{figure}
\begin{center}
\includegraphics[width=5in]{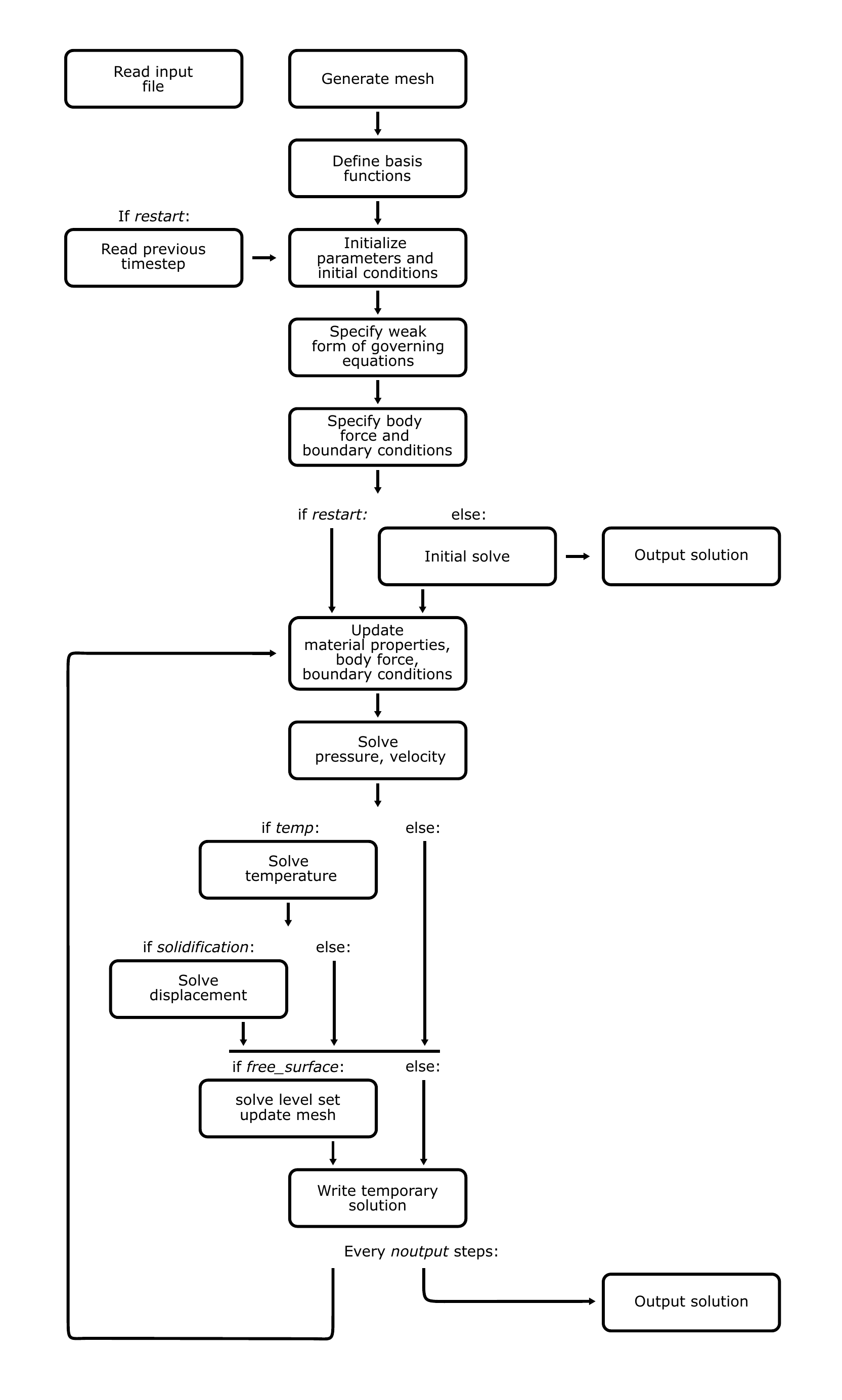}\\
\caption{Schematic flowchart showing each step in the VENUSS algorithm with flags for different model options. }
\label{fig:flowchart}
\end{center}
\end{figure}

We use a split scheme in which the pressure and velocity fields are computed using the previous temperature-dependent rheology and level sets, and then the temperature, displacement, and level sets are updated according to the new velocity field. This reduces the effect of extreme shear localization in response to the strongly non-linear temperature dependence of viscosity and is appropriate for the case of advection-dominated heat transfer (high $\mathrm{Pe}$ limit). We allow for a user-prescribed number of additional iterations through the scheme to update the temperature and level set locations, and therefore the material properties, and forcing in response to the projected temperature field and location of the interface. A schematic of the algorithm steps is shown in Fig. \ref{fig:flowchart}. \par

\subsection{Inf-sup condition and stabilization}
The inf-sup or Ladyzhenskaya–Babuška–Brezzi (LBB) condition gives the criteria for stability of the discrete solution to the Stokes equations \citep{Babuska1973,Brezzi1974}. We satisfy the inf-sup condition for Stokes flow by selection of basis functions that use Taylor-Hood quadrilateral elements (Q2-Q1). This method is stable for incompressible flows, and \citet{Kellogg1996} demonstrate stability for compressible Stokes flow when the pressure field is continuous. We provide additional stabilization options using either (1) Streamline upwind Petrov-Galerkin (SUPG) or (2) Galerkin Least squares (GLS) on the pressure field for compressible flows, free-surface level set, elastic displacement, and temperature. 

\subsection{Assembly and solver}
The problem is assembled using the GetFEM C++ library \citep{Renard2020} through the Python interface. GetFEM uses the Gmm++ for linear algebra and uses its own version of \textit{SuperLU} 3.0. The problem is specified using the GetFEM Generic Weak Form Language and assembly is automated. The non-linear solver uses the Newton-Raphson iterative method. Residuals from the iterative solve are output to the standard error and run until the residual is below a threshold value (e.g., 1\e{-12}) or the maximum number of iterations (e.g., 10) is reached. Maximum convergence is typically reached within approximately three iterations. If the tolerance is not met, the program will continue to the next step, potentially producing spurious results.

\section{Model verification}
We consider a series of test problems to demonstrate the capability to solve multi-physics problems. Initially, we choose simplified problems to isolate the various constituent components required to model lava flows and compare against manufactured or analytical solutions when possible. Then we combine multiple physical processes and demonstrate the effect of different assumptions on model results that highlight the ability of the model to simulate problems of geologic interest. \par

We first solve the lid-driven cavity problem with a manufactured exact solution to demonstrate convergence of pressure and velocity (Section \ref{sec:lid-driven_cavity}). Next, we solve for the relaxation of a free surface with an initial sinusoid compared to a linearized analytical solution to assess convergence of the level set solution and the effect of the free-surface stabilization parameters (Section \ref{sec:free-surface}). Then, we model a domain cooled from one side to show convergence of the temperature solution under diffusion-dominated heat transport to an analytical solution (Section \ref{sec:stefan}). In this section we also demonstrate the solution of velocity and displacement for fluids with strongly temperature-dependent viscosity and solidification in one direction with analytical solutions. Together, these examples cover the dominant processes of lava transport: viscous flow, elastic deformation, movement of a free-surface, and heat transport. 

To demonstrate the relevance to geologic processes, we simulate a semicircular droplet with a cooled shell fed by a volumetric source from below to observe patterns in displacement: comparing a high-viscosity formulation versus an elastic shell to highlight differences in the expected surface deformation patterns. 

\subsection{Lid-driven cavity flow}
\label{sec:lid-driven_cavity}
\subsubsection{Problem description}
\begin{figure}
\begin{center}
\includegraphics[width=\textwidth]{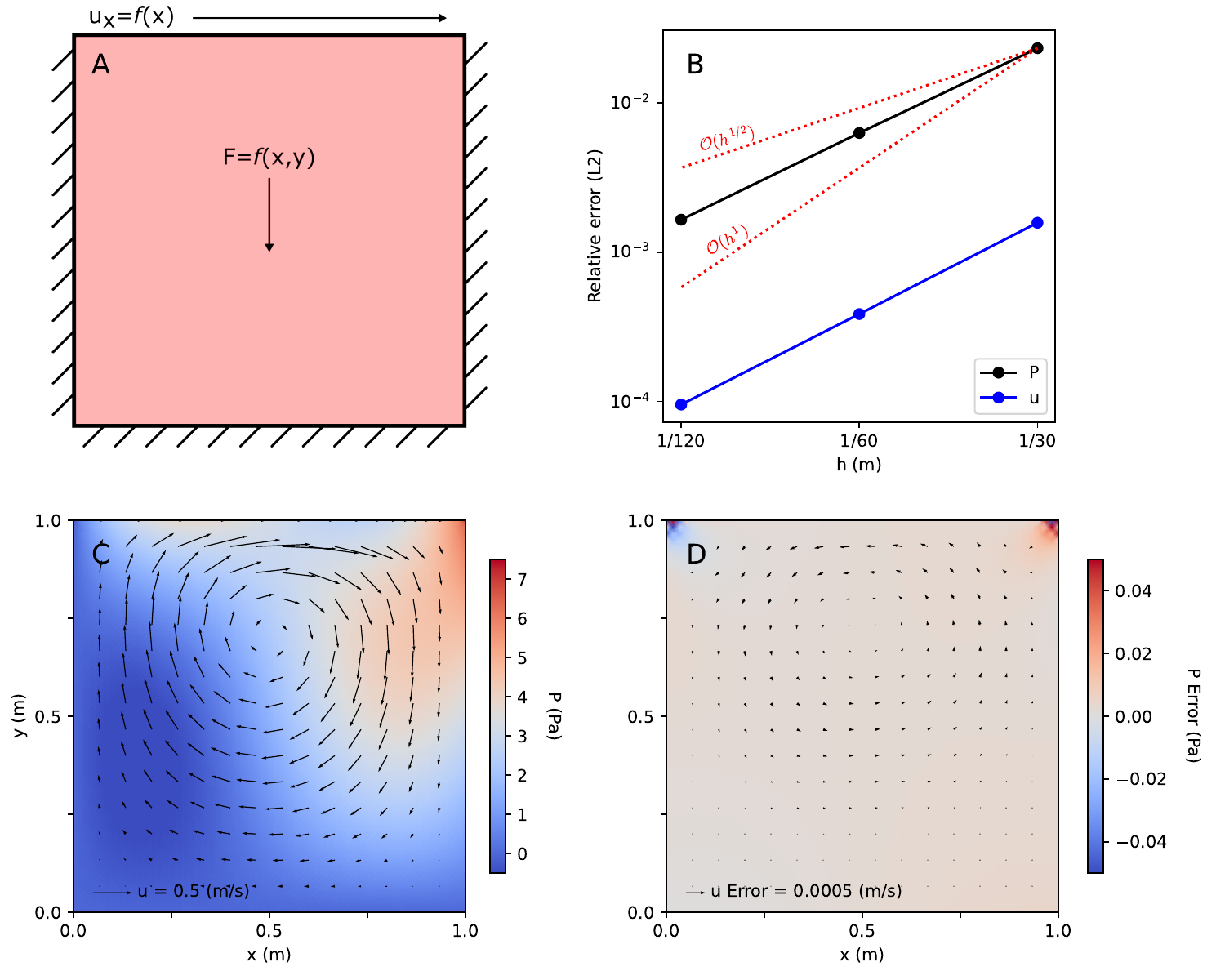}\\
\caption{A) Domain and boundary conditions for test problem of lid-driven cavity flow. B) Convergence (L$_2$ norm) in pressure (black) and velocity (blue) with one-half and first order convergence (dashed red). Example pressure (color) and velocity (vectors) solution (C) and error (D). }
\label{fig:cavity_combined}
\end{center}
\end{figure}

We begin with verification of the incompressible Poisson solution for pressure and velocity in lid-driven cavity flow, illustrated in Fig \ref{fig:cavity_combined}A. We use the problem definition of \citet{Shih1989} in the low $\mathrm{Re}$ limit in which: 
\begin{subequations}
\begin{align}
        \Div \mathbf{u} &= 0 \: , \\
        \Grad^2 \mathbf{u} - \Grad P + \mathbf{f} &= 0 \: , 
\end{align} 
\end{subequations}
where $\mathbf{f}$ is a body force. The boundary conditions are chosen to be: 
\begin{equation}
    \begin{cases}
        u_x = 0, u_y = 0 \:  & \text{on } x = 0 \: ,  \\
        u_x = 0, u_y = 0 \:  & \text{on } x = L_x \: ,  \\
        u_x = 0, u_y = 0 \:  & \text{on } y = 0 \: , \\
        u_x = 16(x^4 -2x^3 + x^2), u_y = 0 \:  & \text{on } y = L_y  \: , \\
        P = 0 \: & \text{on } x = 0, y = 0 \: , 
    \end{cases}
\end{equation}
where the subscripts $x$ and $y$ indicate the component in the x- and y-directions, respectively. The body force, $f$, is chosen to be:  
\begin{subequations}
\begin{align}
    f_x &= 0 \: , \\
    f_y &= -8\left[ 24\left(\frac{x^5}{5} - \frac{x^4}{2} + \frac{x^3}{3}\right)
    + 2 \left( 4x^3 - 6x^2 +2x \right) \left( 12y^2 - 2 \right)
    + \left( 24x-12 \right) \left( y^4 - y^2 \right) \right] \: .
\end{align}
\end{subequations}
This choice of conditions eliminates the singularities at the upper corners of the cavity in the traditional lid-driven cavity problem when forced at a constant velocity, and yields an exact solution for velocity: 
\begin{subequations}
    \begin{align}
        u_x(x,y) &= 8(x^4 -2x^3 + x^2)(4y^3 - 2y) \: , \\
        u_y(x,y) &= -8(4x^3 - 6x^2 + 2x)(y^4 - y^2) \: , 
    \end{align}
\end{subequations}
and pressure (for unit viscosity, $\mu$=1): 
\begin{equation}
    P(x,y) = 8 \left[ \left(\frac{x^5}{5} - \frac{x^4}{2} + \frac{x^3}{3}\right)
    \left( 24 y \right) + \left( 4x^3-6x^2+2x \right) \left(4y^3-2y\right) \right] \: .
\end{equation}

\subsubsection{Pressure and velocity}
Figure \ref{fig:cavity_combined}C shows solutions to the lid-driven cavity problem and the difference from the analytical solution at increasing spatial resolution. We show convergence of the pressure and velocity fields in the L$_2$ norm upon grid refinement for equal-aspect elements with length $h$ (Fig. \ref{fig:cavity_combined}B). We show between one-half and first-order convergence in velocity and pressure on Q2-Q1 elements. 

\subsubsection{Unsteady flow}
In most cases, the inertial component of lava flows is small compared to the viscous (low $\mathrm{Re}$ limit). For highly fluid lavas, it may be important to retain the inertial term. While we still restrict the model to laminar flow, we allow for unsteady viscous flow. The conservation of momentum becomes:  
\begin{equation}
    \rho \partialt{\mathbf{u}} = \Div \sigma - \Grad P + \mathbf{f}, 
\end{equation}
we scale the equation where $x = L\hat{x}$, $y = L\hat{y}$, $\mathbf{u} = U\mathbf{\hat{u}}$, $t = \tau\hat{t}$, $P = \frac{\mu U}{L}\tau\hat{P}$, $\mathbf{f} = f\mathbf{\hat{f}}$, and find:
\begin{equation}
    \rho \frac{U}{\tau} \frac{\partial \hat{u}}{\partial \hat{t}} = \frac{\mu U}{L^2} \Grad^2 \mathbf{\hat{u}} - \frac{\mu U}{L^2} + f \mathbf{\hat{f}} \: . 
\end{equation}

\begin{figure}
\begin{center}
\includegraphics[width=\textwidth]{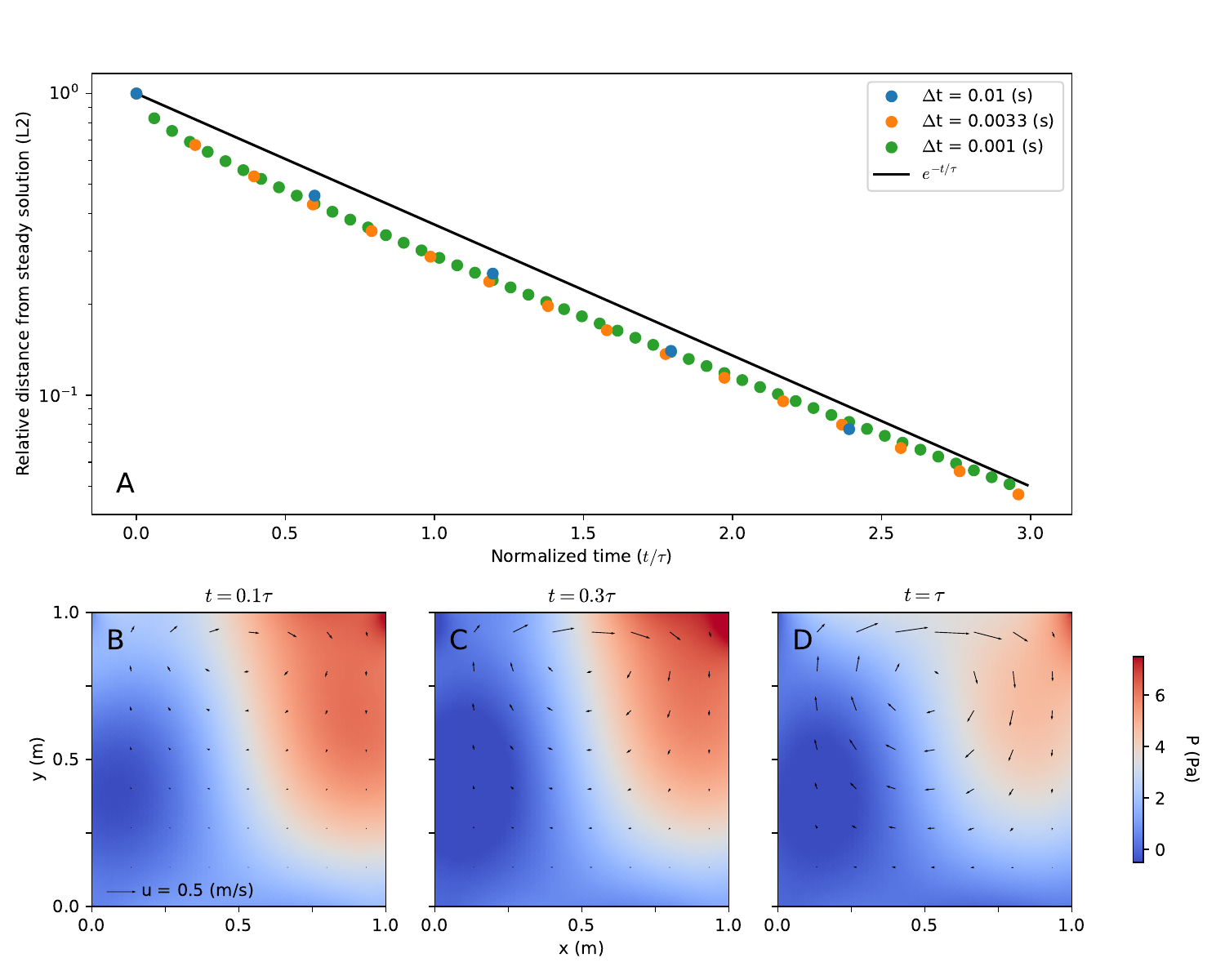}\\
\caption{Approach to the steady-state solution of the lid-driven cavity problem, scaled by the problem timescale $\tau\sim1/33$ s. A) Relative distance from the steady solution for time steps of 0.01 (blue), 0.0033 (orange), and 0.001 (green), and B-D) example snapshots of solutions for pressure (color) and velocity (vectors).}
\label{fig:unsteady}
\end{center}
\end{figure}

Given the problem geometry, $L=1$, $\rho=1$, $\mu=1$, $U=1$, and $f=59.8$. We rearrange to find the problem timescale $\tau\sim0.0167$ s. To demonstrate the model's capability to simulate unsteady flow, we show the approach to the steady-state solution from a fluid at rest, normalized by the timescale of the problem (Fig. \ref{fig:unsteady}). 

\subsubsection{Compressible flow}
\begin{figure}
\begin{center}
\includegraphics[width=\textwidth]{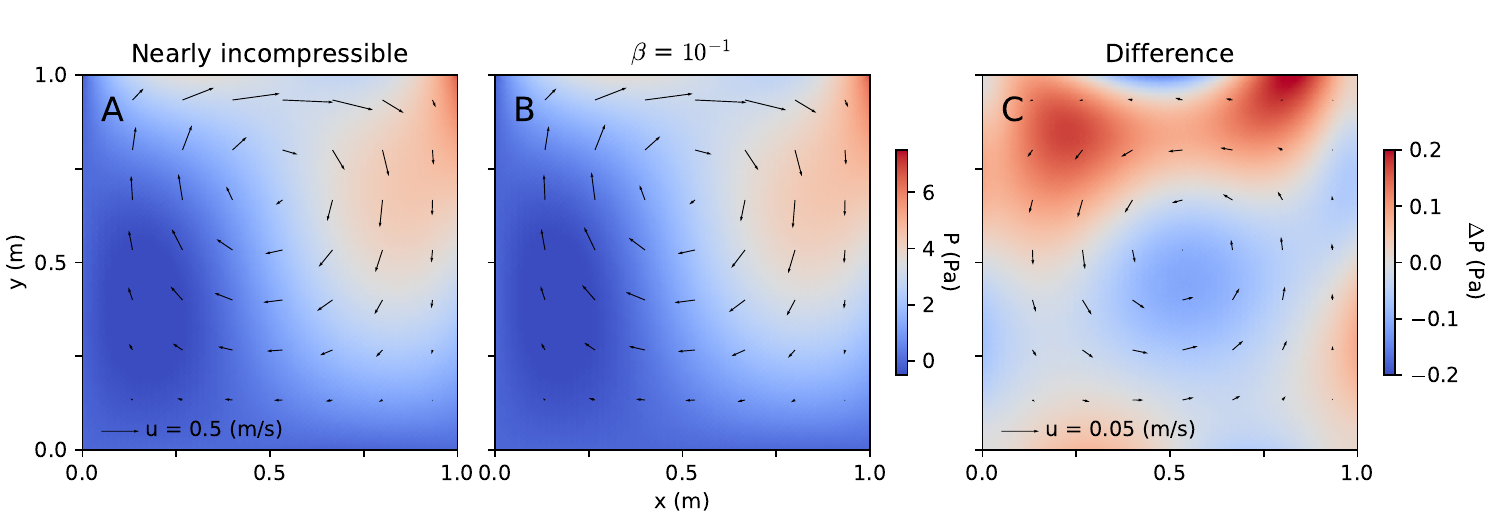}\\
\caption{Steady-state solution of the lid-driven cavity problem for the A) nearly incompressible and B) linearly compressible solutions, and C) the difference (A-B).  }
\label{fig:compressible}
\end{center}
\end{figure}

Although lava flow models can often neglect the effects of compressibility, it may be important for some geometries such as dikes \citep{Rivalta2008}, in high vesicularity flows, and critically for the development of pressure inside a confining coherent crust. To this end, we provide solutions for both the nearly incompressible and compressible cases. We demonstrate the solution by comparing the nearly-incompressible and linear compressible flow in the lid-driven cavity geometry (Fig. \ref{fig:compressible}). 

\subsection{Relaxation of a free surface}
\label{sec:free-surface}
\subsubsection{Problem description}

The evolution of an evolving interface between two fluids with a strong contrast in material properties (a free surface) is of fundamental importance to understanding the flow of lava. Lava and magma have densities (750-3000 kg/m$^3$) and viscosities (50-10$^{12}$ Pas) that are much higher than air ($\rho \sim$1 kg/m$^3$, $\mu \sim 10^{-5}$ Pas) or water ($\rho \sim$1000 kg/m$^3$, $\mu \sim 10^{-3}$ Pas). Additionally, the shape, location, and propagation speed of the free surface of lava flows is often 1) the primary interest of lava flow and dome monitoring for hazard evaluation, and 2) the first-order observable during effusive eruptions. Accurate determination of flow surface location is therefore a main goal of any lava flow simulation. \par

\begin{figure}
\begin{center}
\includegraphics[width=\textwidth]{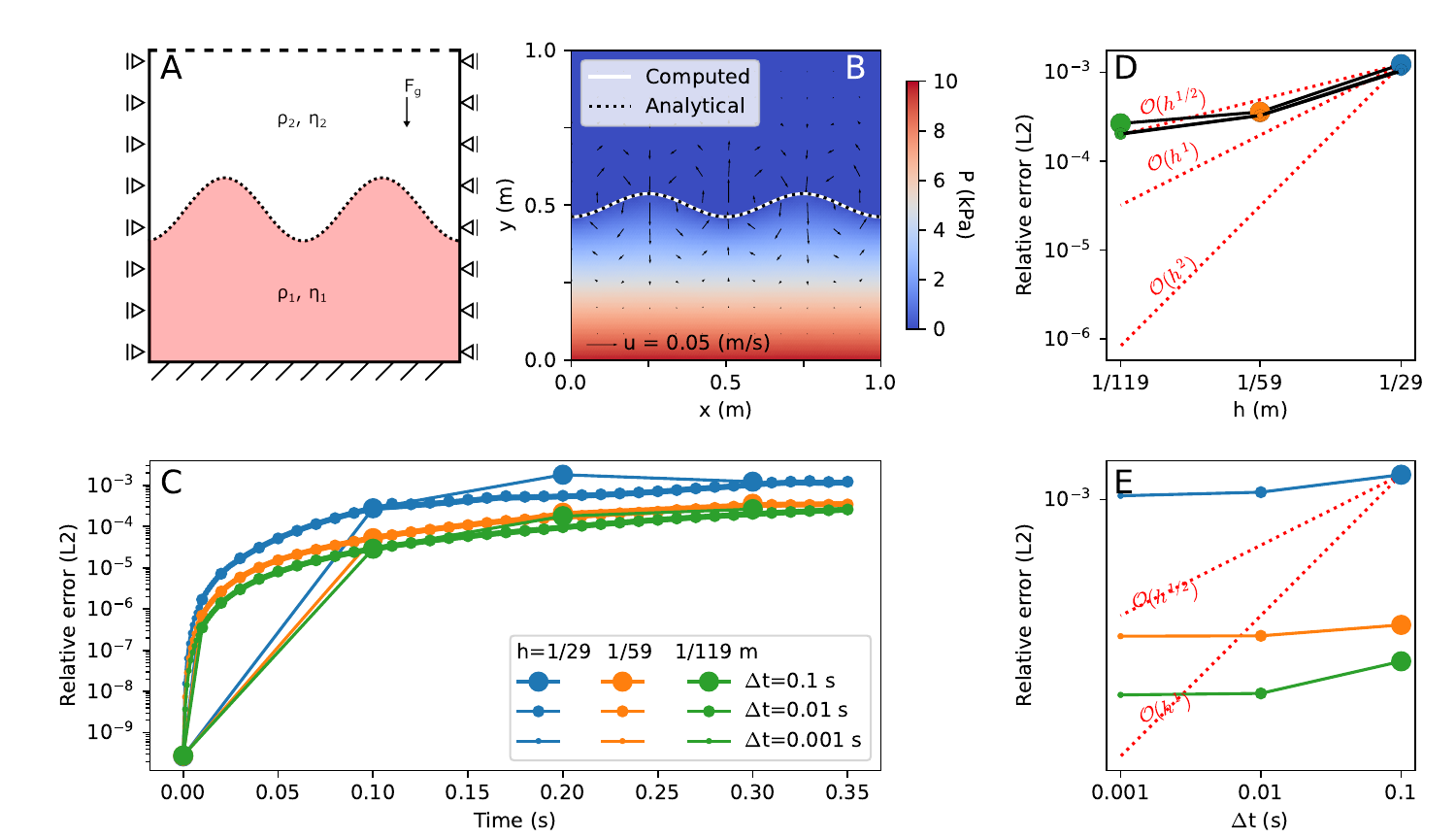}\\
\caption{A) Domain and boundary conditions for test problem of a relaxing free surface with initial sinusoidal interface. B) Example solution at t=0.2 s for pressure (color) and velocity (vectors) with the computed (white) and analytical (dashed black) free surface locations. Relative error (L$_2$ norm) through time (D) and convergence in space (C) and time (E).}
\label{fig:relax_combined}
\end{center}
\end{figure}

We verify the evolution of a relaxing free surface with an initial sinusoidal profile (Fig. \ref{fig:relax_combined}A). We use a square domain with edges of length 1 m, with two fluids. The lower fluid (fluid 1) has a density of 1000 kg/m$^3$ and a viscosity of 200 Pas. The upper fluid (fluid 2) has a density of 1 kg/m$^3$ and a viscosity of 1 Pas. The two fluids have an initial interface of depth, $h_0$:
\begin{equation}
    h_0 (x) = D - A\cos \left( \frac{2 \pi n x} {L_x} \right) \: ,
\end{equation}
defined by a sinusoid with a wavenumber, $n$ = 2, amplitude, $A$ = 0.05 m, centered at depth, $D$ = 0.5 m. \par
Boundary conditions are free-stress on the top, no-slip on the bottom, and roller on the left and right: 
\begin{equation}
    \begin{cases}
        u_x = 0, \partialx{u_y} = 0 \:  & \text{on } x = 0 \: ,  \\
        u_x = 0, \partialx{u_y} = 0 \:  & \text{on } x = L_x \: ,  \\
        u_x = 0, u_y = 0 \:  & \text{on } y = 0 \: , \\
       \partialy{u_x} = 0, \partialy{u_y} = 0 \: & \text{on } y = L_y  \: , \\
       P = 0 \: & \text{on } x = 0, y = L_y \: , 
    \end{cases}
\end{equation}

The problem has a linearized analytical solution for waves with amplitudes that are small compared to the fluid depth and wavelength \citep{Rose2017}: 
\begin{subequations}
    \begin{align}
        h(x,t) &= h_0 (x) \exp \left( -t/\tau \right) \: , \\
        \tau &= \frac{Dk + \sinh\left( Dk \right) \cosh \left( Dk \right)}{\sinh^2\left( Dk \right)} \frac{2 k \eta}{\rho g} \: , \\
        k &= \frac{2 \pi n}{L_x} \: .
    \end{align}
\end{subequations}
This choice of initial condition and material properties has the advantage of using realistic lava properties and length scales and also results in a characteristic timescale of $\tau\approx1$ s. 

\subsubsection{Free-surface location}
We show convergence of the free-surface location with increasing resolution in space and time (Fig. \ref{fig:relax_combined}D\&E) between the computed and analytical solutions. The convergence through time of the solution increases from an initially very low error (discretization of the initial location) to accumulate a relative error of the order of (10$^{-7}$-10$^{-5}$). At long times the simulation trends towards a flat interface, with the Heaviside jump in density allowing for sub-grid resolution of the force acting on the interface.  The convergence order in space is between $\mathrm{O}(\Delta x^{1/2-2})$ and in time is initially first order, but at short time steps is dominated by the error introduced by the spatial discretization. There is a further possible discrepancy between the computed and linearized analytical solution because the finite amplitude of the initial sinusoid allows for lateral flow of material in the computed solution that is neglected in the linearized analytical solution, which results in the true (and computed) solution being slightly under-relaxed compared to the analytical solution.

\subsubsection{Free-surface stabilization parameters}
\begin{figure}
\begin{center}
\includegraphics[width=\textwidth]{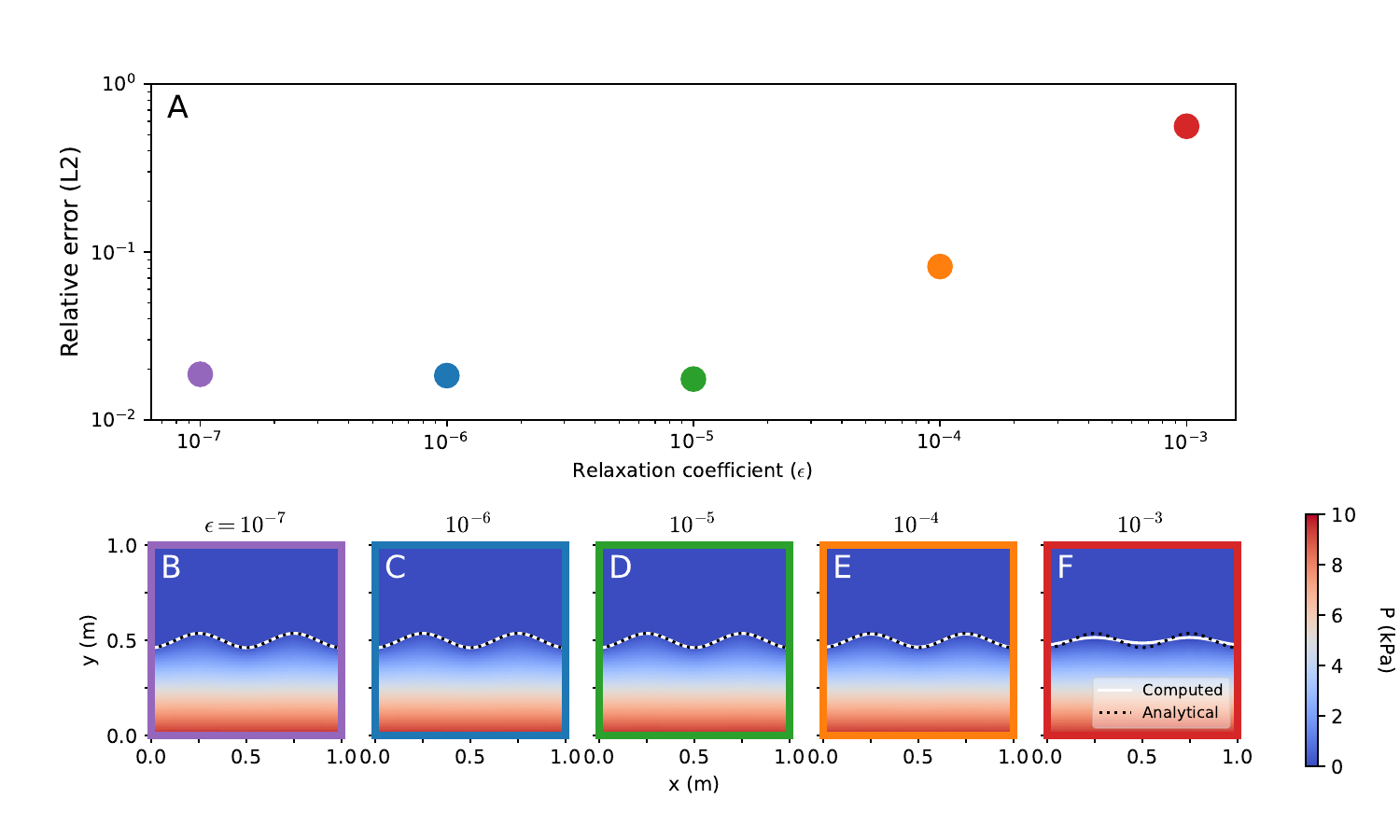}\\
\caption{(A) Error after 0.3 s simulation time at different values of the relaxation coefficient, $\epsilon$, that show a local minimum at a value of 0.01. (B-F) the pressure field, computed (white contour) and analytical (dashed black contour) locations of the free surface.}
\label{fig:relaxation-coefficient-snapshot}
\end{center}
\end{figure}

As described in section \ref{section:level-sets}, we impose two stabilization parameters $\epsilon$ and $\kappa_{\Psi}$, which help prevent unreasonable surface roughness and prevent ``swallowtail'' solutions \citep{Sethian1997}. Ideally, the solution should be relatively insensitive to the choice of these parameters, other than that they are small and non-zero. We test values of $\epsilon$ between 10$^{-7}$ to 10$^{-3}$, and find little sensitivity to values for $\epsilon\lesssim 10^{-5}$ (Fig. \ref{fig:relaxation-coefficient-snapshot}), but significant over-smoothing for higher values of $\epsilon$. Additional tests show that a value of $\epsilon=10^{-5}$ performs well across different spatial and temporal discretizations, and when varying the viscosity or density of the material within a factor of two. We expect the range of tested conditions should cover most lava- and magma-relevant use cases, but this parameter can be tuned by the user, which may be necessary if the material contrasts are significantly higher. \par 

\begin{figure}
\begin{center}
\includegraphics[width=\textwidth]{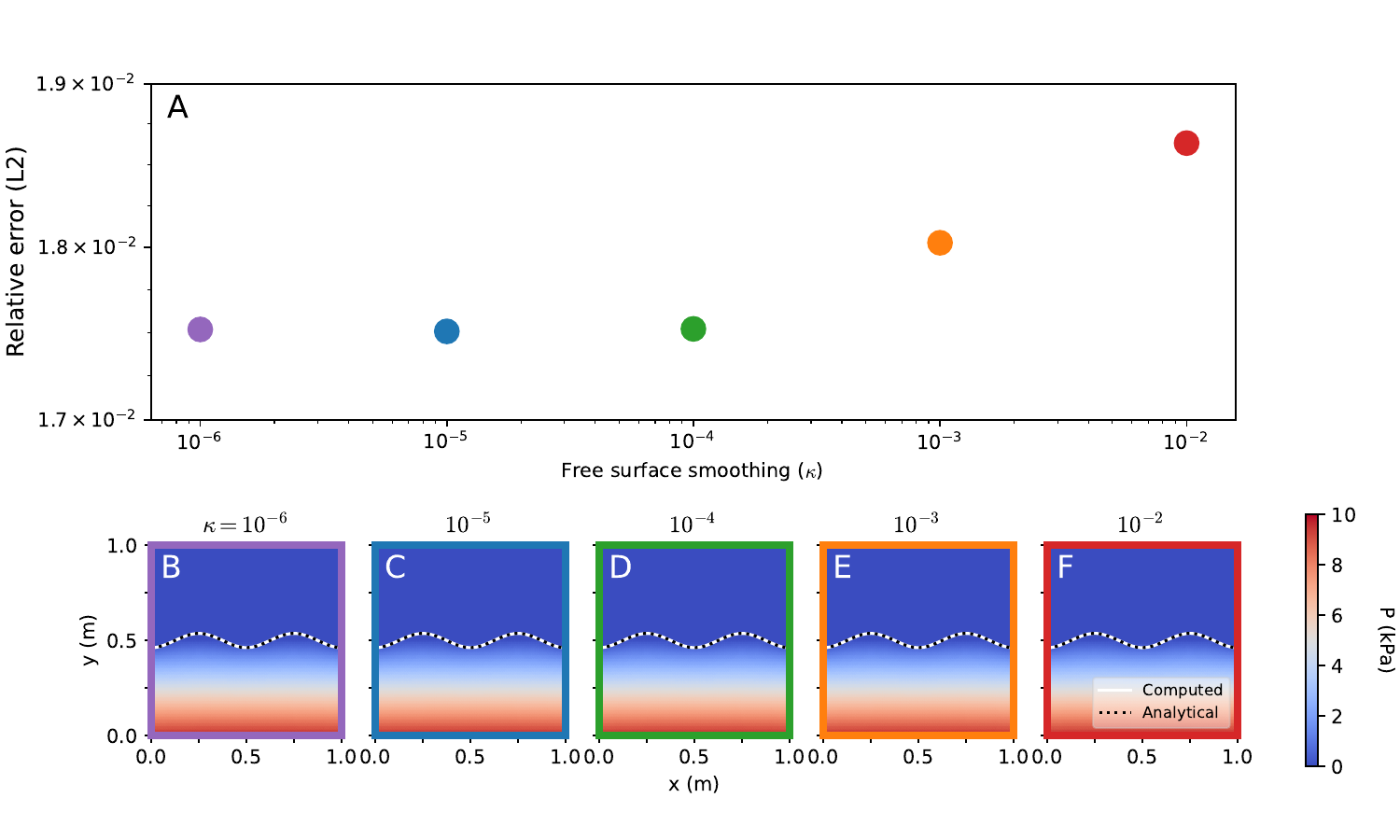}\\
\caption{(A) Error after 0.5 s simulation time at different values of the level-set smoothing parameter, $\kappa_\Psi$, that show little sensitivity over five orders of magnitude. (B-F) the pressure field, computed (white contour) and analytical (dashed black contour) locations of the free surface.}
\label{fig:diffusivity-snapshot}
\end{center}
\end{figure}

The role of stabilization parameter, $\kappa_\Psi$, is to remove high-frequency noise in the level set function that can lead to large estimates of the curvature and result in instability in the smoothing of the interface. In the problem setup (eq. \ref{eq:stabilized_velocity}), we scale $\kappa_\Psi$ by the viscosity of the problem to allow higher-viscosity materials maintain a rougher interface compared to low viscosity materials without requiring case-by-case tuning. The artificial diffusion smooths wavelengths of order $L \lesssim \sqrt{\kappa_\Psi}$. For this test problem that has a characteristic wavelength of the sinusoid $L=0.25$ m, we expect significant smoothing of the problem wavelength when $\kappa_\Psi\sim 0.0625$. The solution is insensitive to the value of $\kappa_\Psi$ over several orders of magnitude (Fig. \ref{fig:diffusivity-snapshot}), suggesting that a small amount of smoothing is adequate for good estimates of curvature in smooth problems. We find good behavior for values of $\kappa_\Psi \sim 10^{-4}$-$10^{-6}$. \par  

\subsection{Solidification}
\label{sec:stefan}
\subsubsection{Problem description}

\begin{figure}
\begin{center}
\includegraphics[width=\textwidth]{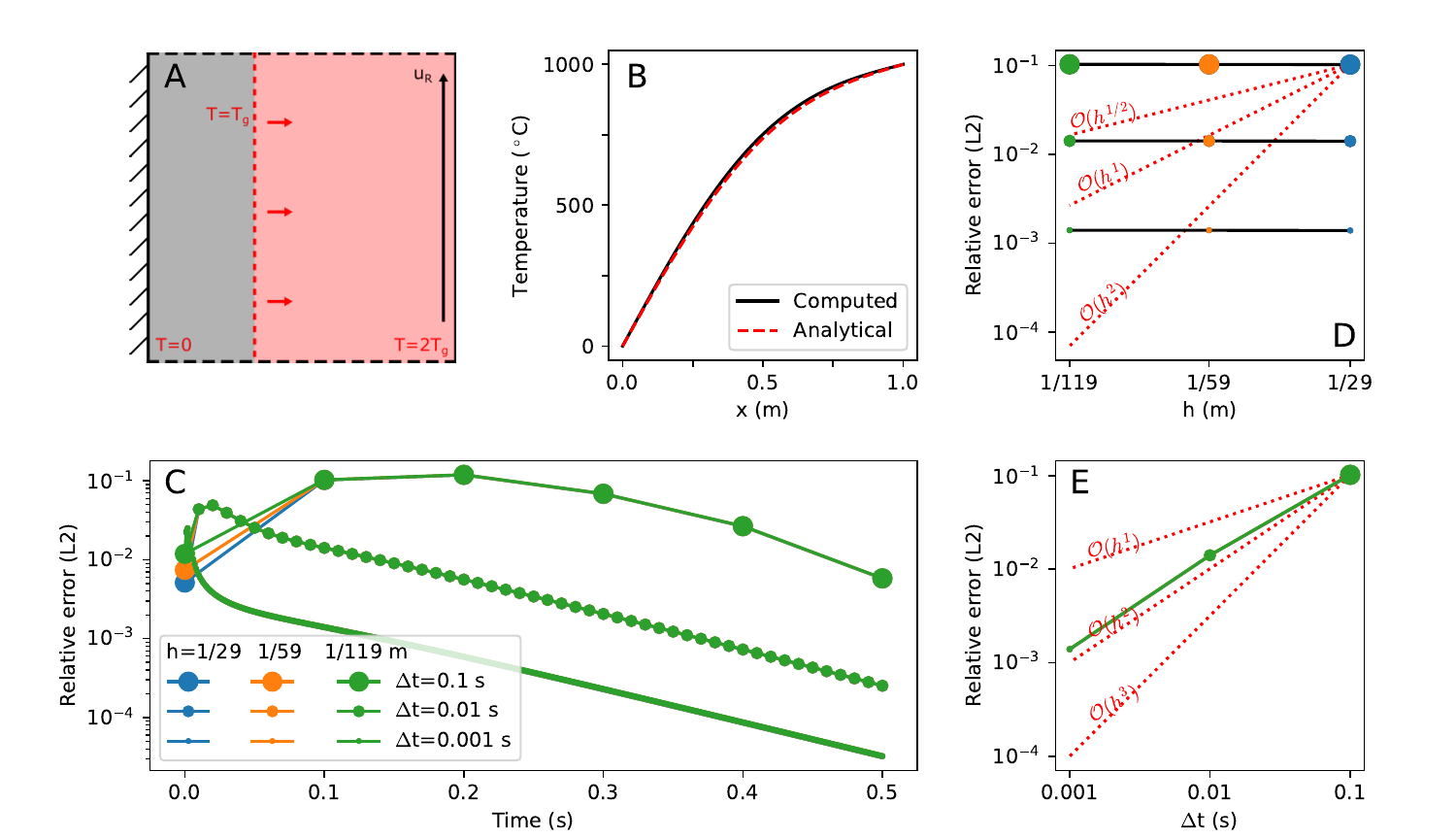}\\
\caption{A) Domain and boundary conditions for test problem of a solidifying fluid, fixed on the left and forced flow on the right. The interface between elastic and viscous domains is determined by the glass transition temperature, here T$_g$=500 $^\circ$C. B) Temperature solution (black) and analytical solution (red dashed). Relative error (L$_2$ norm) through time (D) and convergence in space (C) and time (E).)}
\label{fig:solidification_combined}
\end{center}
\end{figure}

The self-consistent formulation of the transition between viscous and elastic regions in a cooling fluid is the primary contribution of this model. We present tests of the solidification problem to demonstrate this unique capability; we begin with a simple, pseudo-one dimensional problem to verify that we can accurately capture this process. We begin with a viscous fluid cooling in one-direction under simple shear (Fig. \ref{fig:solidification_combined}A) which allows for analytical solutions to the viscous and elastic problems. The domain begins at a constant temperature: 
\begin{equation}
    T(x,y,0) = T_0 \: ,
\end{equation}
and cools under diffusion, which is the primary mechanism in which thick crusts in lavas grow. While forced convection in the surrounding fluid and radiation are also important mechanisms, these terms appear as a sink or source term at the flow surface which is here set to zero. The full boundary conditions are: 
\begin{equation}
    \begin{cases}
        u_x = 0, u_y = u_L, T = 0 \:  & \text{on } x = 0 \: ,  \\
        u_x = 0, u_y = u_R, T = T_0 \:  & \text{on } x = L_x \: ,  \\
        u_x = 0, \partial{u_y} = 0, \partialy{T} = 0 \:  & \text{on } y = 0 \: , \\
        u_x = 0, \partial{u_y} = 0, \partialy{T} = 0 \: & \text{on } y = L_y  \: , \\
        P = 0 \: & \text{on } x = 0, y = 0 \: , 
    \end{cases}
\end{equation}
where we choose $u_L$=0 m/s, $u_R$=0.001 m/s, and $T_0$=1000 $^\circ$C. This geometry has analytical solutions for velocity, pressure, and temperature. The pressure field everywhere is: 
\begin{equation}
    P(x,y,t) = 0 \: , 
\end{equation}
and the velocity field is:
\begin{subequations}
    \begin{align}
        u_x(x,y,t) &= 0 \: , \\
        u_y(x,y,t) &= u_L + \frac{u_R-u_L}{\int_0^{L_x}\frac{1}{\eta(T(x,y,t))}dx} \int_0^x\frac{1}{\eta(T(\hat{x},y,t))} d\hat{x} \: .
    \end{align}
\label{eq:stefan_noniso}
\end{subequations}
The temperature field has an analytical solution (Fig. \ref{fig:solidification_combined}B): 
\begin{subequations}
    \begin{align}
    \label{eq:stefan_t_sol}
        T(x,y,t) &= T_0 \frac{x}{L_x} +  \sum_{n=1}^{\infty} T_0 \exp \left( -K_T k_n^2 t \right) (-k_n^2) \sin \left( k_nx \right) \: , \\
        k_n &= \frac{n \pi}{L_x} \: .
    \end{align}
\end{subequations}

\subsubsection{Diffusion-dominated temperature}
We show convergence of temperature in the diffusion-dominated problem between the computed solution and the analytical solution. The evolution of the error over time (Fig. \ref{fig:solidification_combined}C) shows that most of this error occurs in the early part of the simulation, during which the higher frequency modes are damped out. Because the time step is coarse with respect to the spatial discretization, this effect depends primarily on the time step and is insensitive to the spatial discretization (Fig. \ref{fig:solidification_combined}D\&E). After this point, the error decreases through time as the computed and analytical solutions both lose the high-frequency component due to diffusion and evolve to a linear profile in space. We find first-order convergence with time (Fig. \ref{fig:solidification_combined}E), but all are overly diffusive compared to the analytical solution, which is expected using the stabilized (either SUPG or GLS) Galerkin formulation, and given the sharp nature of the initial condition. 

\subsubsection{Temperature-dependent viscosity}
\begin{figure}
\begin{center}
    \includegraphics[width=\textwidth]{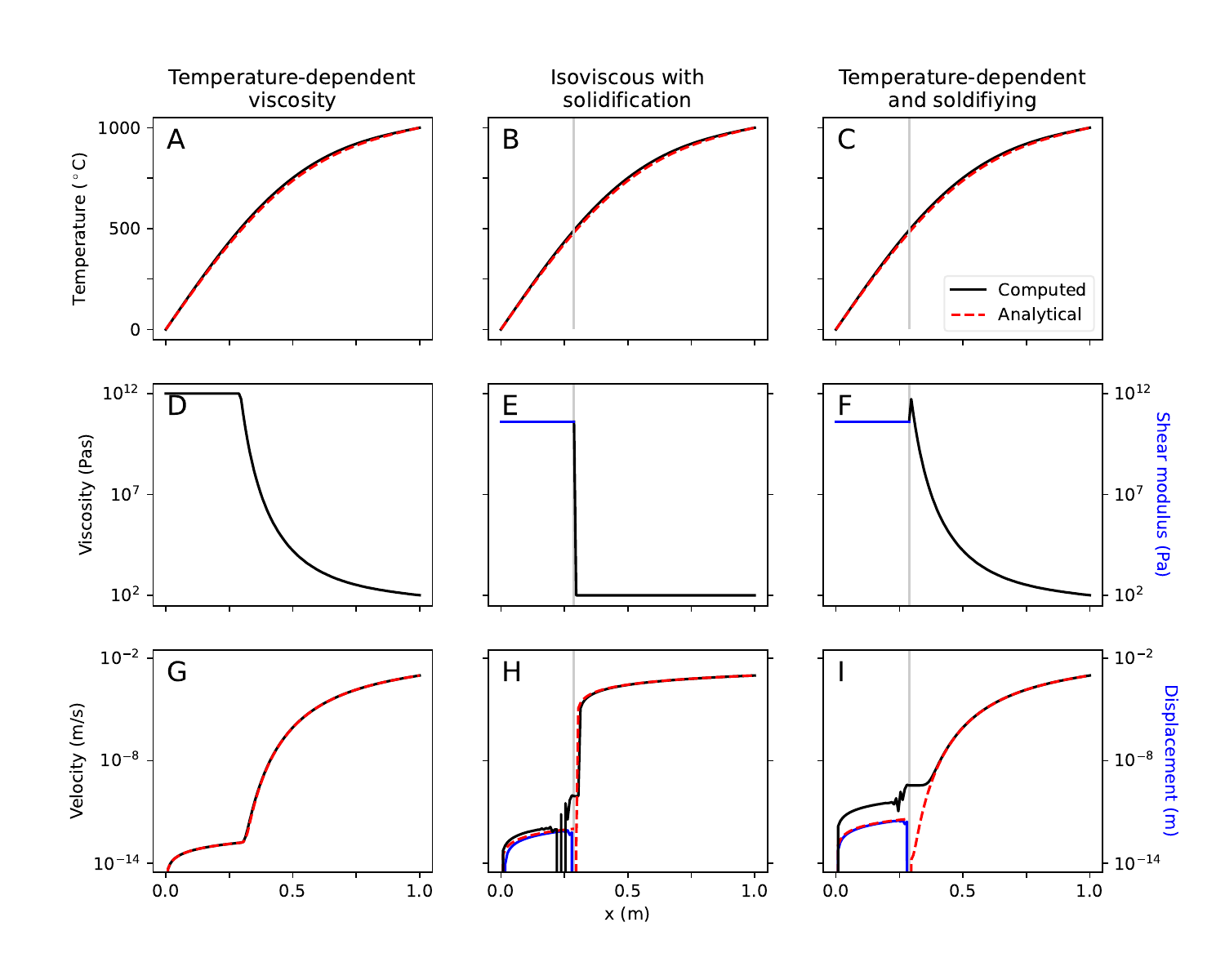}
    \caption{Temperature solution along the centerline (black, A-C), viscosity (black, D-F), shear modulus (blue, E-F), velocity (black, G-I), and displacement (blue, H-I) with gray vertical lines indicating the solidification front and the analytical solutions to each field (dashed red A-C \& G-I) for fluids with temperature-dependent viscosity without solidification (A,D\&G), isoviscous with solidification (B,E\&H), and temperature-dependent viscosity with solidification (C,F\&I).}
\label{fig:stefan_solid_combined}
\end{center}
\end{figure}

In natural lavas and magmas, the viscosity varies over many orders of magnitude in response to temperature changes prior to solidification. We demonstrate the capability of the model to handle strongly temperature-dependent viscosity described by the VFT equation. We choose $A$ = -2, $B$ = 2800, and $C$=300 $^\circ C$, as they cover a range between 100 Pas at 1000 $^\circ$C and the maximum viscosity of $\eta_{\text{max}}=$1\e{12} Pas (glass transition) at $T_g$ = 500 $^\circ$C for numerical convenience, but this spans the range of temperatures and viscosities for typical basaltic eruptions and shares a shape (fragility) similar to that of lava. We find good agreement between the computed velocity and a numerical solution to Eq. \ref{eq:stefan_noniso} using the trapezoidal rule and the computed viscosity (Fig. \ref{fig:stefan_solid_combined}G). \par

\subsubsection{Solidifying front}
We add to the problem, a moving internal boundary between a high-temperature, viscous part of the domain, and an elastic part of the domain, with the boundary defined on $T=T_g$ (a Stefan problem). Given the temperature solution in Eq. \ref{eq:stefan_t_sol} and the solidification front location, $s(t)$ has a short-term solution of the form $s_{S}\sim \sqrt (\kappa_T t)$ and approaches at long times $s_L = L_x/2$. There are inherently matching stress, and no-slip conditions on the internal boundary due to the unified problem description and smooth basis functions in velocity and displacement. If the timescale of the elastic and viscous response is short with respect to the timescale of cooling, the velocity solution is: 
\begin{equation}
u_x = 0, u_y =
    \begin{cases}
        u_L & \text{if } x\leq s(t) \: , \\
        u_L + \frac{u_R-u_L}{\int_{s(t)}^{L_x}\frac{1}{\eta(T(\hat{x},t))}d\hat{x}} \int_{s(t)}^x\frac{1}{\eta(T(\hat{x},t))} d\hat{x} & \text{if } x>s(t)
    \end{cases}
\end{equation}
and the elastic displacement: 
\begin{equation}
    d_x = 0, d_y = \begin{cases}
    d_L + \frac{u_R-u_L}{\int_{s(t)}^{L_x}\frac{1}{\eta(T(\hat{x},t))}d\hat{x}} \int_{0}^{x}\frac{1}{G(T(\hat{x},t))} d\hat{x}  & \text{defined where } x\leq s(t)
    \end{cases} \: .
\end{equation}
We make the simplifying assumption that the shear modulus is constant, which reduces to: 
\begin{equation}
    d_x = 0, d_y = \begin{cases}
    d_L + \frac{u_R-u_L}{\int_{s(t)}^{L_x}\frac{1}{\eta(T(\hat{x},t))}d\hat{x}} \frac{x}{G}  & \text{defined where } x\leq s(t)
    \end{cases} \: .
\end{equation}
In the iso-viscous case, this has the exact solution for velocity: 
\begin{equation}
    u_x = 0, u_y = 
    \begin{cases}
        u_L & \text{if } x<=s(y,t) \: , \\
        u_L + (u_R-u_L) \frac{x - s(t)}{L_x - s(t)} & \text{if } x>s(t)
    \end{cases}
\end{equation}
and elastic displacement: 
\begin{equation}
    d_x = 0, d_y = \begin{cases}
    d_L + \frac{u_R-u_L}{L_x-s(t)} \frac{\eta}{G} x & \text{defined where } x\leq s(t)
    \end{cases} \: .
\end{equation}
We show results for both isoviscous and temperature-dependent viscosities with solidified regions (Fig. \ref{fig:stefan_solid_combined}H\&I, respectively). In general, the displacements are at or lower than expected from the analytical solution, because the temperature front evolves on the same timescale as the elastic response when we neglect the propagation of elastic waves, resulting in transient non-zero velocities within the solidified region (black curves in Fig. \ref{fig:stefan_solid_combined}H\&I). \par 

\section{Model validation}
\subsection{Dome deformation}
\begin{figure}
\begin{center}
\includegraphics[width=\textwidth]{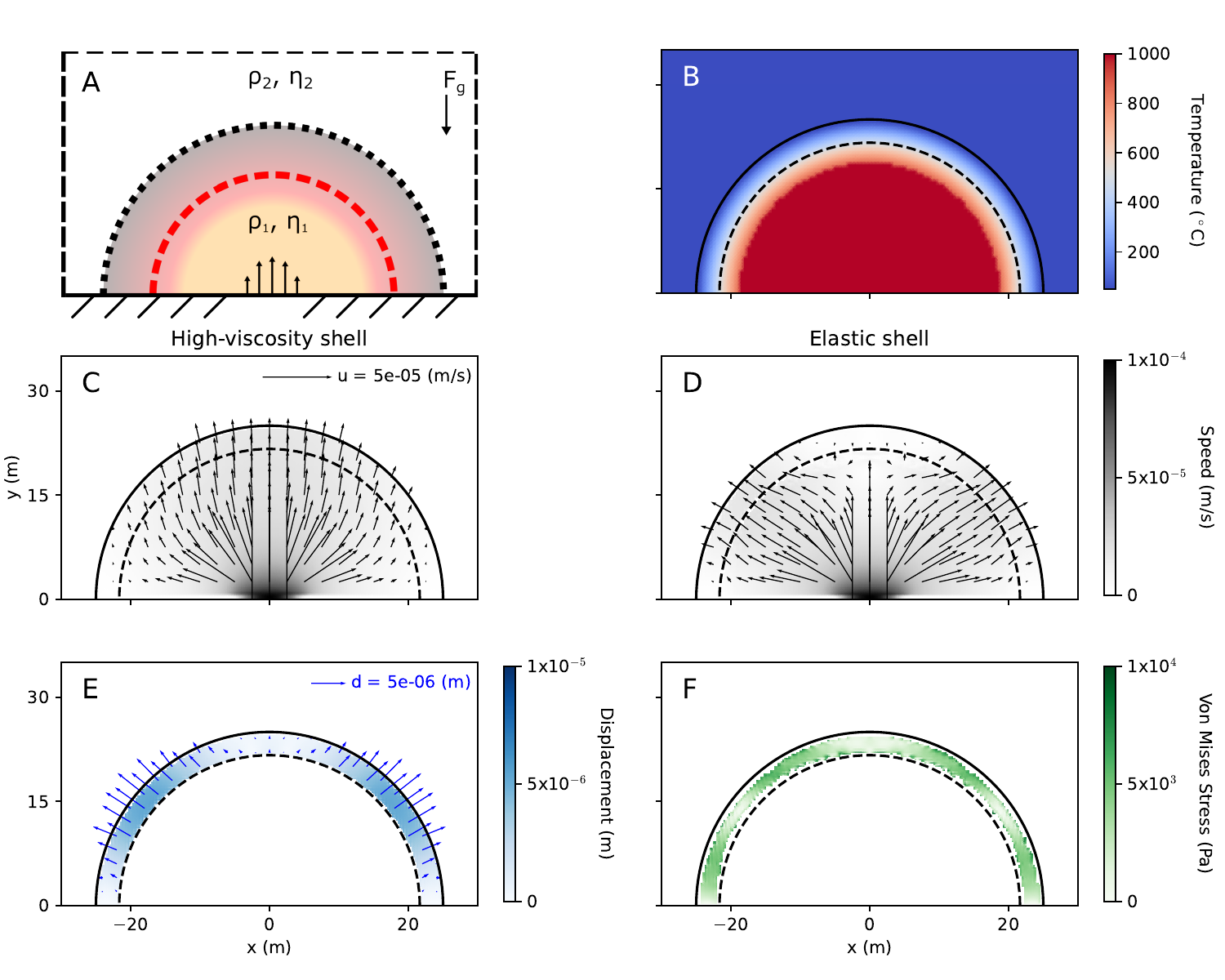}\\
\caption{A) Domain and boundary conditions for a dome-like geometry fed from below by a fixed velocity, B) initial temperature field with with solid line indicating the free surface and dashed line the isotherm corresponding to the solidification front. Comparison of velocity in the C) purely viscous case, where the viscosity above the "solid" isotherm has a constant high viscosity, and the D) elastic case. E) Displacement, and F) von Mises stresses in the elastic shell. }
\label{fig:dome_combined}
\end{center}
\end{figure}

We demonstrate the difference in expected deformation between a purely viscous (although temperature dependent) model and that of an elastic shell. We begin with a hemispherical dome (2D semicircle within a planar framework) (Fig. \ref{fig:dome_combined}A), initialized with a temperature field with a linear gradient from 0 $^\circ$C at the free surface to 800 $^\circ$C over a shell of thickness 6 m, and then increases over a further 1 m to an interior of 1000 $^\circ$C (Fig. \ref{fig:dome_combined}B). This results in a ``solidified'' shell of 3.75 m thickness with either 1) a constant, high viscosity that is 7 orders of magnitude larger than the interior, or 2) an elastic shell with an shear modulus equal to the product of the shell viscosity and the time step to maintain a consistent pressurization across the two simulations. The interior boundary of the shell in both cases has a high-viscosity region which transitions to the high-temperature, low-viscosity interior, using the temperature-dependent viscosity of $A$=2.2, $B$=1960, $C$=300 $^\circ$C. We force stresses in the crust by applying a source of material from a 10 m wide conduit at the bottom by setting the bottom Dirichlet boundary condition in velocity to have a parabolic upward velocity with a maximum of 1\e{-4} m/s. Away from the conduit, the bottom has a no-slip condition. \par 

In a simulation that is purely viscous, we observe material moving upwards with vertical velocities smoothly decreasing in magnitude as material spreads laterally outwards from the source (Fig. \ref{fig:dome_combined}C). The maximum velocity within the shell is directly above the source and oriented vertically, with an $\approx$10\% decrease between the shell interior and the surface. In contrast, the simulation with an elastic shell under the same starting conditions and forcing, shows maximum velocity oriented $\approx$40$^\circ$ from horizontal, with almost no deformation directly above the source (Fig. \ref{fig:dome_combined}D\&E). The displacement and velocity at this location decrease across the shell by $\approx$5\%. The particulars of the location of maximum displacement are sensitive to the shell thickness and shape, inflow conditions, and the presence of gravitational forcing, but these simulations highlight the difference in deformation between the viscous and elastic assumptions, critical for the interpretation of lava domes in which the surface deformation is often the principal observable. \par 

From the displacements and the elastic material properties, we find the von Mises stress in two-dimensions: 
\begin{equation}
    \sigma_{\text{VM}} = \sqrt{\sigma_{11}^2 - 2 \sigma_{11} \sigma_{22} + \sigma_{22}^2 + 3 \sigma_{12}^2} \: , 
\end{equation}
where $\sigma_{11}$ and $\sigma_{22}$ are the normal stresses in the x and y directions respectively, and $\sigma_{12}$ is the shear stress. When the von Mises stress exceeds the tensional strength of the material, it is expected to fail under tension or shear. In the case of this geometry, the highest likelihood of failure occurs at two local maxima at $\approx$15$^\circ$ and $\approx$55$^\circ$ above the horizontal (Fig. \ref{fig:dome_combined}F), where the entire thickness of the shell is in a state of high failure likelihood. \par

\section{Discussion \& Conclusions}
We describe a new finite element model: Viscous-Elastic Numerically Unified Solver for Solidifying flows (VENUSS) that resolves a viscous flow interior coupled to an elastic solid shell for modeling solidifying lava flows and domes. The model performs well in a series of example problems with known solutions and captures the fluid-structure interactions required to model the development of crusts in solidifying lava flows. \par 

Our results demonstrate that wholesale cooling of the entire lava flow or dome depth, or even layered models with high viscosity contrast, are insufficient for modeling the surface deformation under a coherent crust or carapace and at cooling margins or flow fronts that deviate from the long-wave and purely-vertical thermal gradient assumptions. Material at the flow margins shows spatial variability in temperature and resultant mechanical properties. The velocities in this region depart from a simple parabolic profile (with no-slip at the bottom to a maximum at the flow surface) predicted by the long-wavelength approximation. Crucially, we find that the inclusion of elastic behavior results in different deformation patterns than a high-viscosity region in dome-like geometries. This suggests that existing models that use an average temperature to increase viscosity, even those specifically targeted at lava domes, may not be able to capture the evolution of dome surfaces or flow fronts. \par

An essential goal of the current work is the improved understanding of the likelihood of brittle failure of the solidified crust of lava flows and domes. With this formulation, we can recover the elastic stresses from the strain field and material properties for calculation of the likelihood of failure. While the model does not yet incorporate a mechanism for the initiation and propagation of fractures, we lay a groundwork from which failure modeling could be accomplished through, for example, phase field or discrete crack modeling, which would expand our understanding of what happens after cracks form and could help identify the conditions required to allow continued or renewed propagation. However, the addition of discrete cracks introduces extreme shear localization and feedbacks which 1) are highly sensitive to small perturbations in the shell thickness and curvature and 2) focus deformation to scales which are not resolved by modeling of the entire dome or flow front, which limits the predictive power of simplified, process-based models to make quantitative predictions about flow evolution. Instead, a distributed plastic failure may be a more appropriate choice for modeling at large scale, and VENUSS allows for the prediction of these regions. The underlying architecture incorporates a simple flag for the solidified region, which is currently a function of only temperature, but is readily extensible to a stress- or history-dependent failure condition that allows for both distributed brittle failure accommodated across a fracture network, and fracture healing in the absence of continued deformation, which we leave for future work.  \par 

\section{Author contributions}
JB: conceptualization, methodology, software, formal analysis, writing - original draft, visualization; EL: conceptualization, methodology, writing - review \& editing, supervision, funding acquisition; MWS: methodology, writing - review \& editing, supervision; JEK: writing - review \& editing, funding acquisition, YL: writing - review \& editing, funding acquisition.

\section{Competing interests}
The authors have no competing interests to declare. 

\section{Data \& code availability}
All software and input files to generate data are available at \url{https://zenodo.org/uploads/19220847} and \url{https://github.com/JanineBirnbaum18/VENUSS}.

\bibliography{references}

\end{document}